\documentclass[usenatbib]{mnras}

\usepackage{newtxtext}
\usepackage[varvw]{newtxmath}  
\usepackage{graphicx}          
\usepackage{microtype}
\hypersetup{linkcolor=red,hypertexnames=false}     

\usepackage[capitalize, nameinlink, noabbrev]{cleveref} 

\newcommand{\dd}{\mathrm{d}}

\title[General spectral Jeans solver]{A general spectral solver for the axisymmetric Jeans equations:\\ fast dynamical modelling of galaxies with arbitrary anisotropy}
    
\author[M.~Cappellari]{Michele Cappellari\thanks{E-mail: michele.cappellari@physics.ox.ac.uk}\\
Sub-Department of Astrophysics, Department of Physics, University of Oxford, Denys Wilkinson Building, Keble Road, Oxford, OX1 3RH, UK}

\date{Accepted 2026 February 26. Received 2026 February 17; in original form 2026 January 22}

\pagerange{\pageref{firstpage}--\pageref{lastpage}} \pubyear{2026}

\begin{document}

\maketitle
\label{firstpage}

\begin{abstract}
Axisymmetric Jeans modelling is widely used to infer galaxy mass profiles from integral-field kinematics, but existing implementations maintain tractability by adopting highly restricted anisotropy prescriptions. I present a new spectral method that solves the axisymmetric Jeans equations as a two-dimensional boundary-value problem. Remarkably, this breaks the traditional trade-off between model flexibility and computational cost, accommodating completely general anisotropy distributions $\beta(r,\theta)$ while executing significantly faster than standard restrictive techniques. The method relies on three key choices: (i) solving for the intrinsic dispersion $\overline{v_r^2}$ rather than the rapidly varying pressure $\nu\overline{v_r^2}$ to improve numerical conditioning; (ii) working in logarithmic radius to efficiently resolve the large dynamic range of galaxies, uniquely matching scale-free (power-law) regimes; and (iii) imposing a Robin outer boundary condition that enforces the correct asymptotic decay on a finite computational domain. Orbit integrations in realistic galaxy potentials motivate spherical alignment of the velocity ellipsoid as a physically plausible default, though the framework easily adapts to other alignments. Validated against exact analytic benchmarks---including new analytic Jeans solutions derived herein---the solver recovers intrinsic second moments with high accuracy, showing radially uniform residuals for power-law tests. In practice, it delivers orders-of-magnitude speed-ups over high-accuracy quadrature schemes and is naturally suited to massive GPU parallelization. Released in the public \textsc{JamPy} package, this enables the routine application of highly general Jeans models to large surveys and the extensive parameter-space exploration required for rigorous uncertainty quantification.
\end{abstract}


\section{Introduction}
\label{sec:intro}

\subsection{Dynamical modelling in the era of large surveys}

Dynamical models of galaxies are the fundamental scales on which the Universe is weighed. By mapping the gravitational potential, these models allow one to measure the masses of supermassive black holes---essential for understanding feedback processes and galaxy co-evolution---and provide the most direct method to infer the distribution of dark matter by separating it from the luminous stellar component. This decomposition is critical for testing cosmological models and reconstructing the assembly history of galaxies.

The two most widely used techniques for these tasks are orbit-superposition methods \citep{Schwarzschild1979} and solutions based on the Jeans equations \citep{Jeans1922}. These approaches are highly complementary, with the choice between them often dictated by the quality and quantity of the available data. Schwarzschild models provide a fully general treatment of phase-space structure, but they require high Signal-to-Noise ($\mathcal{S/N}$) spectra to reliably extract the full Line-of-Sight Velocity Distribution (LOSVD) needed to exploit that flexibility. In contrast, Jeans modelling relies only on the lowest-order velocity moments (mean velocity and dispersion). These moments can be measured reliably even from data with moderate $\mathcal{S/N}$ ratios or at high redshift where the full LOSVD is inaccessible. This robustness makes Jeans modelling particularly well-suited for the massive datasets produced by modern Integral Field Spectroscopic (IFS) surveys such as ATLAS\textsuperscript{3D} \citep{Cappellari2011p1}, CALIFA \citep{Sanchez2012}, SAMI \citep{Croom2012}, and MaNGA \citep{Bundy2015} \citep[see][for reviews]{Cappellari2016,Cappellari2026}. A theoretical caveat is that the moments approach does not guarantee the positivity of the underlying phase-space Distribution Function (DF).

Crucially, the simpler assumptions of the Jeans method do not necessarily result in lower accuracy for mass recovery in realistic situations. \citet[fig.~8]{Leung2018} benchmarked dynamical models against circular velocities derived from CO gas kinematics in 54 galaxies and found that in the outer regions ($0.8\text{--}1.6 R_{\rm e}$, where $R_{\rm e}$ is the projected half-light radius), the measurement errors for the Jeans Anisotropic Multi-Gaussian Expansion (JAM) method were approximately a factor of 1.7 smaller than those for the Schwarzschild method. A similar conclusion was reached by \citet[fig.~4]{Jin2019} using numerical simulations; in recovering the enclosed mass within $R_{\rm e}$, the scatter in the JAM results was a factor of 1.6 smaller than that of the Schwarzschild technique. Neither study found evidence of systematic bias in the JAM results. These findings confirm that, for mass measurement, the restrictive assumptions of Jeans modelling act as a regulariser that stabilizes the solution against observational noise.

Beyond its robustness, Jeans modelling offers superior computational efficiency. Unlike orbit-superposition methods or $N$-body simulations, it avoids the costly integration of thousands of orbits. This speed is not merely a matter of convenience; it enables a level of scientific rigor that is often computationally prohibitive for more complex methods. The efficiency of the Jeans approach allows for the exploration of vast parameter spaces, facilitating the computation of rigorous Bayesian posteriors and the testing of various physical assumptions---such as different dark matter profiles or anisotropy formulations---to ensure results are statistically well-grounded.

\subsection{Past applications of the JAM method}
\label{sec:jam_applications}

A prominent implementation of the Jeans modelling approach is the Jeans Anisotropic Modelling (JAM) method \citep{Cappellari2008,Cappellari2020} and its associated \textsc{JamPy} software package,\footnote{Current version 8.1 at \url{https://pypi.org/project/jampy/}} which is built on the \textsc{Numpy} \citep{Numpy2020} and \textsc{Scipy} \citep{Scipy2020} ecosystem. This framework has established itself as a standard tool for extracting mass distributions across a vast range of scales and redshifts.
It has received widespread adoption, serving as the primary technique for the dynamical modelling of stellar kinematics in major integral-field spectroscopic (IFS) surveys. Notable examples include the modelling of all early-type galaxies in ATLAS\textsuperscript{3D} \citep{Cappellari2013p15}, the SAMI survey \citep{Scott2015}, the CALIFA survey \citep{Leung2018}, and MaNGA \citep{Li2017imf,Li2019}, culminating in the dynamical analysis of $\sim$10,000 galaxies in the MaNGA DynPop project \citep{Zhu2023DynPop1, Lu2023DynPop2}. 

The specific dynamical studies associated with these large samples have enabled robust constraints on key galaxy properties. These include the total density slopes \citep[e.g.][]{Cappellari2015dm, Poci2017, Li2019, Zhu2024DynPop3} and dark matter fractions \citep{Cappellari2013p15,Lu2024DynPop5}, the systematic variation of the stellar initial mass function \citep[IMF;][]{Cappellari2012, Li2017imf, Lu2024DynPop5}, and the origins of the fundamental plane \citep{Cappellari2013p15, Shetty2020, Zhu2024DynPop3}. 

Beyond global structural parameters, these modelling efforts have also been instrumental in exploring the link between dynamical quantities and stellar populations. Notably, the local escape velocity $V_{\rm esc}$, derived from JAM models in the cited studies, was found to correlate tightly with stellar population diagnostics \citep{Scott2009, Scott2013p21}. It was also demonstrated that stellar population properties closely follow lines of constant velocity dispersion $\sigma_{\rm e}$, or equivalently $M_*\propto R_{\rm e}$, on the $(M_*, R_{\rm e})$ stellar mass-size plane \citep{Cappellari2013p20,McDermid2015,Cappellari2016}. In the local Universe, the framework is currently being applied to the GECKOS survey data to study the asymmetries of edge-on discs from high-quality stellar kinematics \citep{Rutherford2025}. 

The method has also been extended to higher redshifts ($z \approx 0.8$--$2$) to trace the evolution of galaxies over cosmic time. Specific dynamical analyses include the measurement of dynamical masses \citep[e.g.][]{Cappellari2009apjl, Mendel2020} and total density profiles \citep{Derkenne2021,Mozumdar2026} using data from surveys like LEGA-C \citep[e.g.][]{vanHoudt2021,Cappellari2023}, KMOS\textsuperscript{3D} \citep{Ubler2024}, and MAGPI \citep[e.g.][]{Derkenne2023}. Most recently, with the advent of the \textit{JWST}, the method has been pushed even further, enabling dynamical modelling out to $z \sim 5$ \citep{Pascalau2026}.

In the context of cosmology, JAM is frequently combined with strong gravitational lensing to break the mass-sheet degeneracy and constrain the total mass profile of lens galaxies \citep{Barnabe2012, Posacki2015, Yildirim2020}. This joint lensing-dynamics approach has been critical for the TDCOSMO collaboration's precise measurements of the Hubble constant $H_0$ \citep[e.g.][]{Shajib2023, Tdcosmo2025} and even for testing general relativity on galactic scales \citep{Collett2018}.

On smaller scales, the method has been adapted to model discrete tracers, providing insights into the dark matter haloes of Milky Way dwarf spheroidal satellites \citep[e.g.][]{Vasiliev2018,Yang2025} and the internal dynamics of Globular Clusters \citep[e.g.][]{Watkins2013}. It has also been used to construct detailed dynamical models of the Milky Way disk using full six-dimensional phase space data from \textit{Gaia} and APOGEE \citep[e.g.][]{Nitschai2020, Nitschai2021} to break the mass-anisotropy degeneracy.

Finally, JAM has been widely employed to determine black hole masses across a broad demographic range in both ETGs \citep[e.g.][]{Krajnovic2018, Shetty2020, Thater2019, Thater2022} and spiral galaxies \citep[e.g.][]{Nguyen2026m81}. This includes landmark benchmark measurements in the elliptical galaxy M87 \citep{Simon2024} and the spiral galaxy NGC~4258 \citep{Drehmer2015,Nguyen2026maser}, which provided critical cross-checks against independent results from Event Horizon Telescope imaging and water maser dynamics, respectively, as well as tests against molecular gas (CO) dynamics \citep{Dominiak2025}. Beyond these massive systems and the nuclear black hole of the Milky Way \citep[sec.~4.1.2]{FeldmeierKrause2017}, the method has pushed into the low-mass regime to detect intermediate-mass black holes in nearby dwarf galaxies and ultra-compact dwarfs \citep[e.g.][]{Seth2014, Ahn2017, Ahn2018, Nguyen2018, Nguyen2019}.

\subsection{The mathematical bottleneck of semi-analytic quadratures}
\label{sec:bottleneck}

The diverse applications of the JAM method described above have historically relied on a mathematical formulation based on semi-analytic quadratures. Established by seminal works on semi-isotropic solutions \citep{Satoh1980, Binney1990, vanDerMarel1991, Emsellem1994} and spherically-aligned anisotropic solutions \citep{Bacon1983, Bacon1985}, this strategy reduces the two-dimensional partial differential equations (PDEs) of the Jeans system to one-dimensional integral representations through judicious changes of variables or the method of characteristics. This approach was further refined into the current JAM method \citep{Cappellari2008, Cappellari2020}.

While this strategy provides accurate solutions, it imposes strict functional limitations on the galaxy model. The Jeans equation for the radial pressure $\nu \overline{v_r^2}$ constitutes a first-order linear ordinary differential equation along specific integration paths. Its general solution requires an \emph{integrating factor} \citep[e.g.][sec.~7.2]{Arfken2013mathematical}, which must be derived analytically to express the final solution as a fast, semi-analytic quadrature. This constraint restricts the anisotropy parameter $\beta$ to specific forms where the integral $\int (\beta/r) \dd r$ is analytic \citep[e.g.][sec.~13.3.1]{Ciotti2021}. The most common choice is a constant $\beta$, which yields a simple power-law integrating factor $r^{2\beta}$, while the most general is the logistic anisotropy of $\log r$ introduced by \citet{Simon2024}. Any deviation from these restricted forms necessitates the numerical evaluation of nested integrals, which destroys the computational efficiency that makes packages like \textsc{JamPy} attractive.

The general Jeans equations are PDEs mathematically analogous to the Euler equations of fluid dynamics; indeed, \citet{Jeans1922} called them the ``hydrodynamical equation of motions of stars.'' One might therefore expect standard fluid-dynamics methods (finite differences or finite elements) to suffice. However, brute-force implementations struggle to deliver both high accuracy and efficiency over the extreme dynamic range of galaxy models, so such approaches have remained niche theoretical tools. I identified only a handful of studies exploring alternative PDE techniques, such as finite differences in spheroidal coordinates \citep{Evans1990} or the method of lines for equations with general velocity-ellipsoid alignment \citep{Yurin2014}; to my knowledge, these have never been used for dynamical modelling of real galaxies.

Furthermore, even spectral methods---the focus of this paper---fail if applied without careful formulation. In my initial attempts using spectral collocation to solve for the standard pressure term $\nu \overline{v_r^2}$, I found that because the stellar tracer density $\nu$ typically declines by many orders of magnitude from the galaxy centre to the outskirts, a polynomial basis struggles to approximate such a rapidly varying function. Using a finite number of terms, these implementations are either insufficiently accurate or require an excessive number of basis functions, leading to a loss of the very efficiency they sought to provide.

\subsection{A new spectral solver}

The main contribution of this paper is a specific reformulation of the Jeans problem that overcomes the mathematical bottlenecks described above. I demonstrate that a spectral method can achieve fast exponential convergence, provided three critical design choices are made:

\begin{enumerate}
    \item \textbf{Solving for the velocity dispersion $\overline{v_r^2}$ rather than the pressure $\nu \overline{v_r^2}$.} As detailed in \cref{sec:reduction}, while the tracer density $\nu$ varies by many orders of magnitude across a galaxy, the velocity dispersion $\overline{v_r^2}$ is a relatively slowly varying function. Solving for this smooth variable improves the conditioning of the problem and allows the spectral representation to be extremely compact, requiring fewer basis functions to achieve high precision.
    \item \textbf{Adopting the logarithmic radial coordinate $x = \ln r$.} As discussed in \cref{sec:mapping}, galaxies span a large dynamic range in radii. The use of logarithmic coordinates ensures that the spectral resolution is distributed uniformly across the logarithmic decades of the galaxy. Moreover, this mapping is mathematically optimal as it transforms the governing radial differential equation with variable coefficients into one with constant coefficients for scale-free (power-law) densities.
    \item \textbf{Imposing a Robin boundary condition on a truncated radial grid.} As discussed in \cref{sec:linear_system}, standard Dirichlet boundary conditions ($\overline{v_r^2}=0$) imposed at a finite radius introduce truncation errors that saturate convergence, preventing the method from reaching machine precision. The adoption of a generalized Robin condition ($r \partial \overline{v_r^2}/\partial r + \mu \overline{v_r^2} = 0$) allows the method to accurately mimic the asymptotic behaviour of the solution even with a compact finite grid.
\end{enumerate}

The practical implications of this reformulation are profound. Using the same modest number of grid points currently employed in \textsc{JamPy} (typically $20 \times 10$), my method achieves higher accuracy than the standard JAM quadrature while recovering the solution at a computational speed up to two orders of magnitude faster. 

This performance leap is accompanied by the removal of the physical constraints inherent to previous integral-based methods. My approach handles completely general anisotropy distributions $\beta(r, \theta)$---for which no analytic integrating factor exists---without any additional computational cost. This speed and flexibility facilitate the exploration of vast parameter spaces, rigorous Bayesian inference, and the systematic testing of different physical assumptions. Furthermore, the method's structure is naturally suited for massive parallelization on GPUs, suggesting that a properly formulated spectral method provides a superior framework for the next generation of dynamical modelling.

The paper is organized as follows. \cref{sec:physics} presents the theoretical framework, justifying the spherical alignment assumption and deriving the master equation for the velocity dispersion. \cref{sec:solver} details the spectral solver algorithm, highlighting the critical role of the Robin boundary condition and the method for spectral interpolation. \cref{sec:mge} provides the implementation formulas for Multi-Gaussian Expansion \citep[MGE;][]{Emsellem1994,Cappellari2002mge} models. \cref{sec:tests} validates the solver against exact analytic solutions and benchmarks its performance. Finally, \cref{sec:conclusions} summarizes the conclusions.

\section{The Spherically Aligned Jeans Framework}
\label{sec:physics}

\subsection{Spherical alignment: the simplest realistic choice}
\label{sec:justification}

\begin{figure*}
    \centering
    \includegraphics[width=0.48\textwidth]{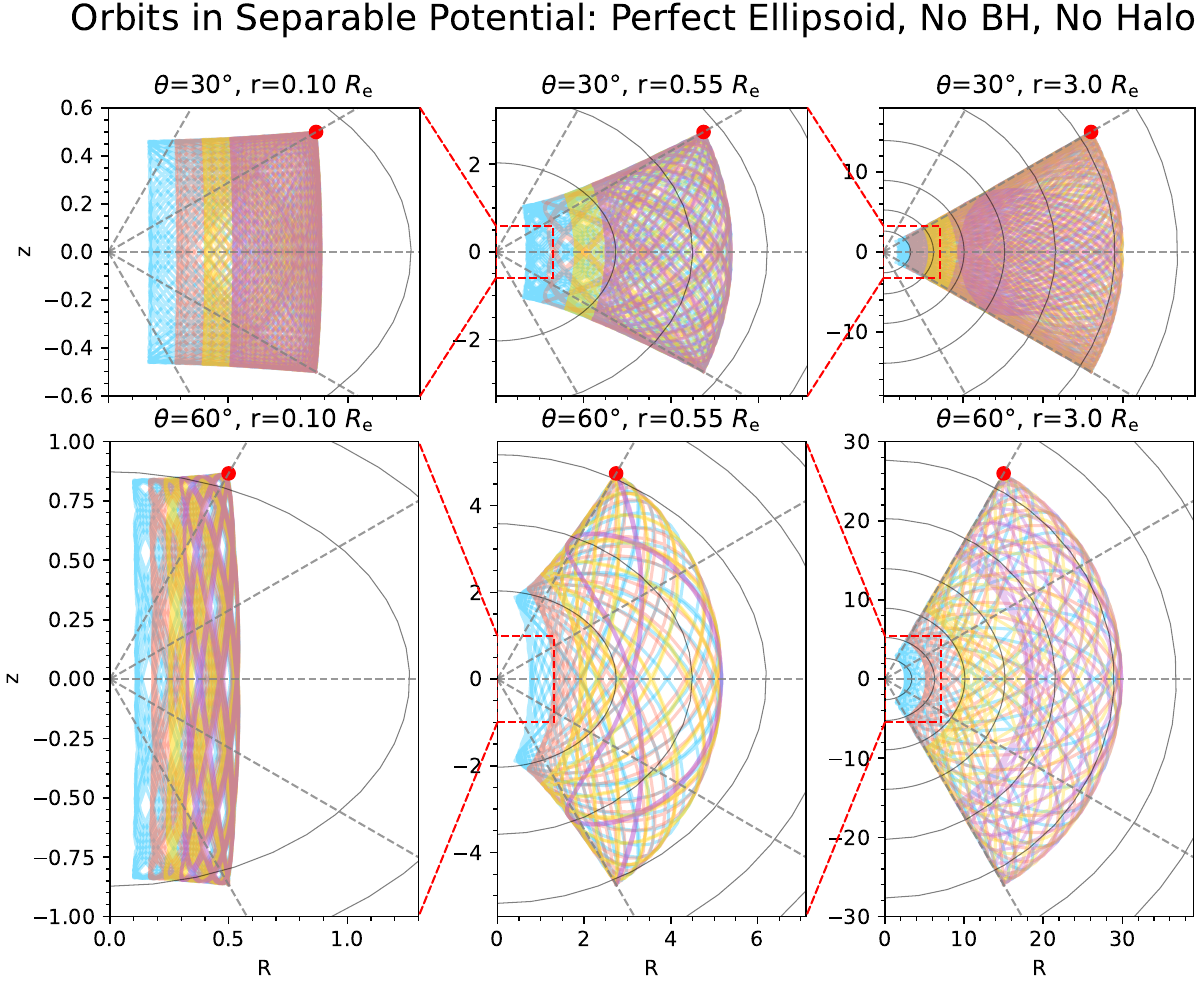}
    \includegraphics[width=0.48\textwidth]{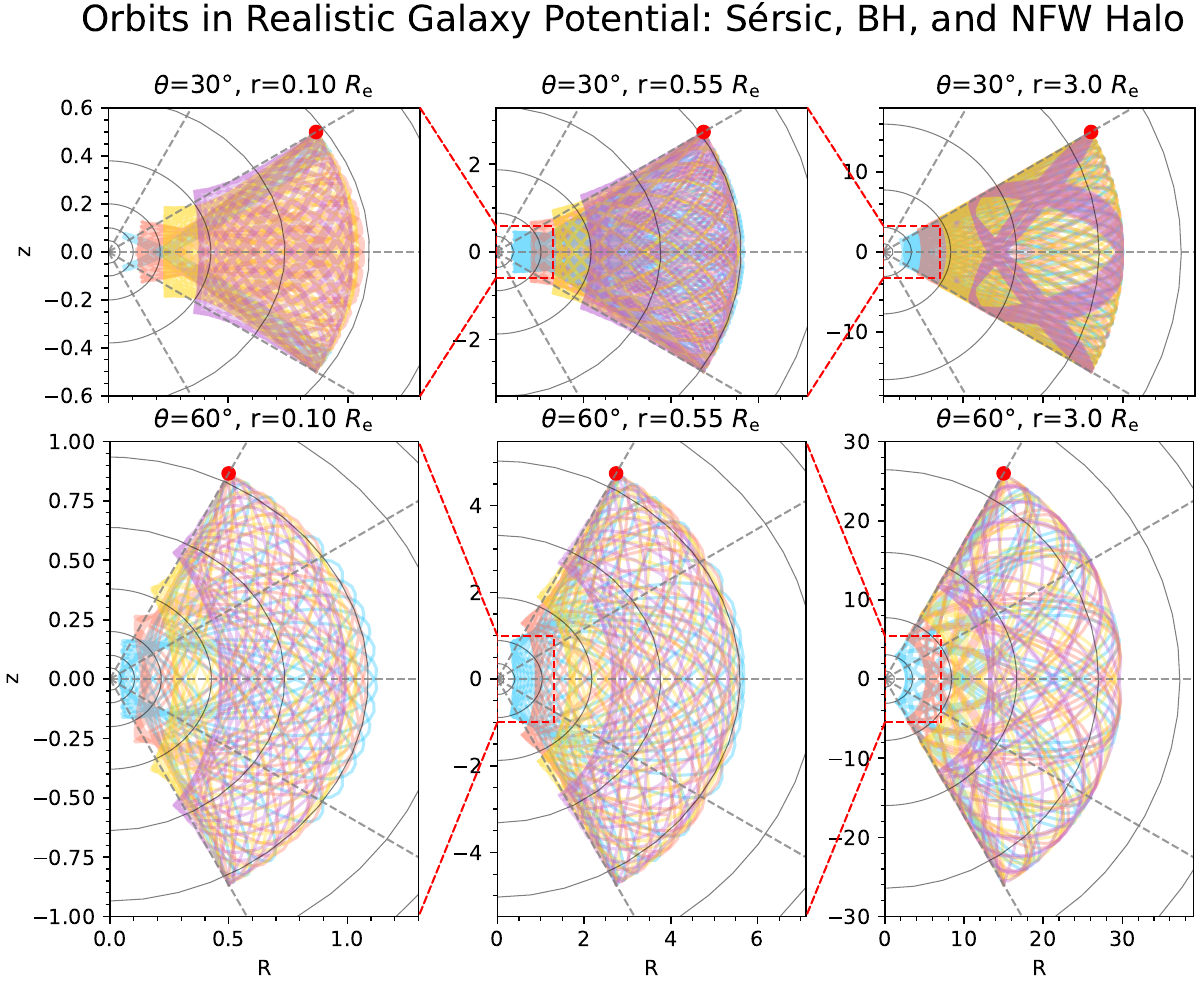}
    \caption{Comparison of orbital envelopes in a separable potential versus a realistic galaxy potential. In each panel, four orbits are launched from the location of the red circle at a distance $r$ from the galaxy centre, with $v_R=v_z=0$ and a range of azimuthal velocities $v_\phi = 0.2v_c(r), \ldots, 0.6v_c(r)$ smaller than the local circular velocity $v_c(r)$. \textbf{Left:} Orbits in a Perfect Ellipsoid potential \citep{deZeeuw1985} with density $\rho(m)=\rho_0/(1+m^2)^2$. Here, $m^2=R^2+z^2/q^2$ is the elliptical radius with an axial ratio $q=0.5$. The orbital boundaries rigidly follow global prolate spheroidal coordinates, forcing the velocity ellipsoid to become cylindrically aligned in the core. \textbf{Right:} Orbits in a realistic multi-component potential, consisting of a \citet{Sersic1968} $n=4$ spheroid with $R_{\rm e}=10$ and $q=0.5$, an NFW halo \citep{Navarro1996nfw} containing $10\%$ dark matter within $R_{\rm e}$, and a Supermassive Black Hole with $M_\bullet = 0.005 M_*$. The gray contours are equipotentials (which are rounder than the density). While the orbits do not strictly follow a single coordinate system, their envelopes are remarkably well approximated by spherical coordinates (dashed lines) at all radii, from the BH-dominated centre ($0.1 R_{\rm e}$) to the halo-dominated outskirts ($3 R_{\rm e}$). This justifies the use of spherically aligned velocity ellipsoids for modelling real galaxies.}
    \label{fig:orbits}
\end{figure*}

To obtain a unique solution of the axisymmetric Jeans equations, one must assume a velocity-ellipsoid orientation. Although historical treatments often adopt cylindrical \citep{Cappellari2008} or prolate spheroidal coordinates \citep{Dejonghe1988, Evans1991, Arnold1995, vandeVen2003}, orbital structure in realistic potentials indicates that spherical alignment \citep{Bacon1983, Bacon1995, Cappellari2020} holds a unique, physically motivated position for describing real galaxies.

The theoretical baseline for velocity ellipsoid alignment was established by \citet{Eddington1915}, who noted that in separable potentials, the principal axes of the velocity ellipsoid align with the coordinate system because the orbital motion decomposes into independent oscillations with zero covariance. This behaviour is strictly realized in Stäckel potentials, such as the Perfect Ellipsoid \citep{deZeeuw1985}. In these systems, orbits rigidly respect a single global coordinate system. As they approach the core, the defining prolate spheroidal coordinates degenerate into cylinders, forcing the velocity ellipsoid into a cylindrical orientation (see \cref{fig:orbits}, left panel).

However, real galaxies are significantly more complex. I examine the orbital behaviour in a realistic, multi-component potential. Unlike the spherical potential used in \cref{sec:sersic_gnfw_test} to allow for an exact analytic benchmark, here the potential is not spherical, but is given by the combination of a flattened ($q=0.5$) \citet{Sersic1968} $n=4$ stellar spheroid, a spherical NFW \citep{Navarro1996nfw} halo, chosen to have a break radius $r_s=10\,R_{\rm e}$ and a dark matter fraction inside a sphere of radius $R_{\rm e}=10$ (in the plot units) of $f_{\rm DM}(<R_{\rm e})=10\%$, typical of massive ETGs (\citealt{Cappellari2013p15}; \citealt{Lu2024DynPop5}; \citealt[fig.~11]{Cappellari2026}), and a supermassive black hole with mass $M_\bullet = 0.005 M_*$ \citep[eq.~11]{Kormendy2013review}. \cref{fig:orbits} (right panel) illustrates the envelopes of regular orbits launched at different inclinations ($\theta=30^\circ$ and $60^\circ$) across three orders of magnitude in radius, from the black-hole dominated nucleus ($0.1 R_{\rm e}$) to the halo-dominated outskirts ($3 R_{\rm e}$).

Two key features emerge from this orbital experiment:
\begin{enumerate}
    \item \textbf{Local vs. Global Coordinates:} While the boundary of any \emph{individual} orbit may exhibit a slight curvature reminiscent of prolate spheroidal coordinates---showing a small `bend' toward the horizontal near the turning points---this effect is local. As is well known, and evident in the zoomed panels of \cref{fig:orbits}, these boundaries do not share a common focal distance. Unlike Stäckel potentials, where a single coordinate system defines all motions, real potentials require a `Stäckel fudge' \citep{Binney2012actions} where the effective coordinate system varies from orbit to orbit. This concept is central to action-based modelling in software like \textsc{GalPy} \citep[e.g.][]{Bovy2015galpy} and \textsc{Agama} \citep{Vasiliev2019agama}.
    \item \textbf{Radial Alignment of Turning Points:} Most crucially, the `corners' of the regular orbits (the extrema where radial and angular velocities vanish) align remarkably well with lines of constant spherical polar angle (dashed grey lines) across all scales. While this alignment is expected in the outer halo or near the central black hole, where the potential becomes spherical, it is not obvious in the intermediate region where the potential is significantly flattened and far from spherical. Yet, as shown, the spherical alignment approximation still applies effectively at all radii.
\end{enumerate}

Consequently, while the velocity ellipsoid of real galaxies cannot be globally described by any simple coordinate system, like spherical, spheroidal, or cylindrical \citep{Evans2016}, spherical alignment represents the most robust global approximation. It captures the fundamental radial nature of the orbital envelopes observed in \cref{fig:orbits} without imposing the artificial constraints that force unphysical cylindrical alignment in the core of separable models. This theoretical expectation is supported by \textit{Gaia} studies \citep{Gaia2016_Mission}, which are the only ones currently able to measure the ellipsoid orientation in a direct manner, from six-dimensional phase space coordinates. These have found that the velocity ellipsoid is well approximated by spherical polar alignment in both the outer stellar halo \citep{Wegg2019} and the disc region \citep{Everall2019, Hagen2019}. See the review by \citet[sec.~3.4]{Hunt2025} for a comprehensive discussion of the observational evidence.

Thus, I adopt spherically aligned velocity ellipsoids not merely for mathematical convenience, but because they provide a superior description of the phase-space structure in realistic, non-integrable galactic potentials.

Finally, I emphasize that the spectral formalism presented here, is not mathematically restricted to the spherically aligned case. Unlike semi-analytic quadrature approaches, which rely on specific integrability conditions for efficiency, the spectral method can be applied with equal ease to the Jeans equations with \textit{any} alignment. This includes cylindrically aligned models \citep[JAM$_{\rm cyl}$;][]{Cappellari2008}, alignment in fixed prolate spheroidal coordinates \citep{Dejonghe1988, Evans1991, Arnold1995, vandeVen2003}, or a velocity ellipsoid with a generic tilt specified by an arbitrary function \citep[e.g.][]{Yurin2014}. One merely needs to rewrite the corresponding PDEs derived in those references, with minimal changes to the implementation. However, given the physical status of spherical alignment discussed above, I consider it the most appropriate standard for general applications and restrict the formulations in this paper to this case.

\subsection{Assumptions and general equations}
\label{sec:assumptions}

I adopt a spherical polar coordinate system $(r, \theta, \phi)$, where the symmetry axis of the galaxy aligns with the $z$-axis ($\theta=0$). The spatial domain covers the radial range $r \in [r_{\min}, r_{\max}]$ and the polar angle $\theta \in [0, \pi/2]$. The dynamics of the tracer population are governed by the Jeans equations under the following three fundamental assumptions:

\begin{enumerate}
    \item \textbf{Steady state:} The system is time-independent ($\partial/\partial t = 0$).
    \item \textbf{Axisymmetry:} The density and potential are symmetric about the $z$-axis ($\partial/\partial \phi = 0$).
    \item \textbf{Spherical alignment:} The velocity ellipsoid is aligned with the spherical coordinate directions. This implies that the off-diagonal components of the velocity dispersion tensor vanish in these coordinates ($\overline{v_r v_\theta} = \overline{v_r v_\phi} = \overline{v_\theta v_\phi} = 0$).
\end{enumerate}

Under these assumptions, the two non-trivial Jeans equations can be written as \citep[e.g.][eq.~2.4]{deZeeuw1996}:
\begin{align}
    \frac{\partial (\nu \overline{v_r^2})}{\partial r} + \frac{2\nu \overline{v_r^2} - \nu \overline{v_\theta^2} - \nu \overline{v_\phi^2}}{r} &= - \nu \frac{\partial\Phi}{\partial r}, \label{eq:jeans_rad_gen} \\
    \frac{\partial (\nu \overline{v_\theta^2})}{\partial \theta} + \frac{\nu \overline{v_\theta^2} - \nu \overline{v_\phi^2}}{\tan\theta} &= - \nu \frac{\partial\Phi}{\partial \theta},
\label{eq:jeans_ang_gen}
\end{align}
where $\nu(r, \theta)$ represents the tracer density and $\Phi(r, \theta)$ denotes the total gravitational potential. 

To close this system, I define the anisotropy parameter as:
\begin{equation}
    \beta(r, \theta) \equiv 1 - \frac{\overline{v_\theta^2}}{\overline{v_r^2}}.
\end{equation}
A key feature of the spectral method presented here is that, unlike traditional quadrature-based approaches, I allow $\beta$ to vary generally as a function of both radius $r$ and polar angle $\theta$. By substituting $\overline{v_\theta^2} = (1-\beta)\overline{v_r^2}$ into \cref{eq:jeans_rad_gen} and \cref{eq:jeans_ang_gen}, the equations become \citep[e.g.][eqs.~1 and 2]{Bacon1983}:
\begin{align}
    \frac{\partial (\nu \overline{v_r^2})}{\partial r} + \frac{(1+\beta)\nu \overline{v_r^2} - \nu \overline{v_\phi^2}}{r} &= - \nu \frac{\partial\Phi}{\partial r}, \label{eq:jeans_rad} \\
    \frac{\partial [(1-\beta)\nu \overline{v_r^2}]}{\partial \theta} + \frac{(1-\beta)\nu \overline{v_r^2} - \nu \overline{v_\phi^2}}{\tan\theta} &= - \nu \frac{\partial\Phi}{\partial \theta}.
\label{eq:jeans_ang}
\end{align}
By solving for the radial and angular components simultaneously, this framework provides the radial $\overline{v_r^2}$ and azimuthal second moment $\overline{v_\phi^2}$ for any given anisotropy and mass distribution. 

As is typical with the Jeans equations, the resulting solution is not guaranteed to be physically realizable; that is, it may not necessarily correspond to a non-negative distribution function. Furthermore, the model should be viewed as an approximation, given that the velocity ellipsoids of real galaxies are never perfectly aligned with a single coordinate system, as argued in \cref{sec:justification}.

\subsection{Solving for the velocity moments in logarithmic coordinates}
\label{sec:reduction}

In traditional Jeans modelling, it is customary to solve for the ``pressure'' term $\nu \overline{v_r^2}$. However, in galactic systems, the tracer density $\nu$ typically spans many orders of magnitude from the nucleus to the outskirts, whereas the velocity dispersion $\overline{v_r^2}$ varies only by a factor of a few. A numerical scheme solving for $\nu \overline{v_r^2}$ must therefore capture an enormous dynamic range, which can lead to a significant loss of precision in the outer regions. To circumvent this, I minimize the dynamic range of the solution by solving directly for the intrinsic second moment $\overline{v_r^2}$. Furthermore, to handle the large radial scales efficiently, I utilize the logarithmic radial coordinate $x = \ln r$.

I begin by eliminating the azimuthal second moment $\nu \overline{v_\phi^2}$ between the radial and angular Jeans equations (\cref{eq:jeans_rad} and \cref{eq:jeans_ang}). By combining these equations and utilizing the vector identity:
\begin{equation}
    r\cos\theta\,\frac{\partial\Phi}{\partial r} - \sin\theta\,\frac{\partial\Phi}{\partial \theta} = r\,\frac{\partial\Phi}{\partial z},
\end{equation}
the right-hand side is reduced to a compact form involving the vertical gradient of the potential, which is computationally efficient to evaluate for several simple models. This procedure generalizes the approach of \citet[eq.~3]{Bacon1983} to allow for a general anisotropy $\beta(r, \theta)$. Throughout the derivation, I multiply the equations by $\cos\theta$ to eliminate the singularity of the $\tan\theta$ term at the equatorial plane ($\theta=\pi/2$), ensuring that all terms remain finite. The resulting partial differential equation (PDE) for the radial pressure $\nu \overline{v_r^2}$ is:
\begin{equation}
-\cos\theta \frac{\partial (\nu \overline{v_r^2})}{\partial \ln r} + \sin\theta \frac{\partial [(1-\beta)\nu \overline{v_r^2}]}{\partial \theta} - 2\beta \cos\theta (\nu \overline{v_r^2}) = \nu r \frac{\partial\Phi}{\partial z}.
\label{eq:bacon_gen}
\end{equation}

To transform this into an equation for the velocity dispersion, I expand the derivatives of the product terms using the product rule and divide the entire equation by the tracer density $\nu$. This yields a linear first-order transport PDE for $\overline{v_r^2}$ with source and reaction terms:
\begin{equation}
-\cos\theta \frac{\partial \overline{v_r^2}}{\partial \ln r} + (1-\beta)\sin\theta \frac{\partial \overline{v_r^2}}{\partial \theta} + \mathcal{C}(r, \theta)\, \overline{v_r^2} = r \frac{\partial\Phi(R, z)}{\partial z}.
\label{eq:final_pde}
\end{equation} 
The coefficient $\mathcal{C}(r, \theta)$ accounts for the spatial logarithmic gradients of the density and angular derivative of the anisotropy:
\begin{equation}
\mathcal{C}(r, \theta) = \sin\theta \left[ (1-\beta)\frac{\partial \ln \nu}{\partial \theta} - \frac{\partial \beta}{\partial \theta} \right] - \cos\theta \left( \frac{\partial \ln \nu}{\partial \ln r} + 2\beta \right).
\label{eq:coefficient}
\end{equation}

This formulation is regular throughout the domain $\theta \in [0, \pi/2]$ and explicitly accommodates angular variations in anisotropy. As discussed in \cref{sec:intro}, $S \equiv \overline{v_r^2}$ is a smooth function, making it ideal for the spectral solver described in \cref{sec:solver}. 

Once the radial dispersion $\overline{v_r^2}$ is obtained, the azimuthal second moment $\overline{v_\phi^2}$ is calculated directly from \cref{eq:jeans_rad}:
\begin{equation}
\overline{v_\phi^2} = \overline{v_r^2} \left( 1 + \beta + \frac{\partial \ln \nu}{\partial \ln r} \right) + \frac{\partial \overline{v_r^2}}{\partial \ln r} + \frac{\partial \Phi}{\partial \ln r}.
\label{eq:v2phi}
\end{equation}
The $\partial \overline{v_r^2}/\partial \ln r$ derivatives, are evaluated spectrally from the previous spectral $\overline{v_r^2}$ solution, to maintain high precision and efficiency.

\section{Spectral Solution Algorithm}
\label{sec:solver}

I solve the governing \cref{eq:final_pde} using a spectral method. While the conceptual idea is simple---approximating the solution with global polynomials that satisfy the PDE at a grid of collocation points---realizing the method's full potential requires a careful study of its technical nuances. A vast technical literature has been developed over the years to ensure the exponential convergence that makes these methods powerful. A clear, practically oriented book on this method is \citet{Trefethen2000}. A brief overview of the key concepts is also presented in \citet[Chapter~20.7]{Press2007}, while more general and comprehensive books on spectral methods are available by \citet{Boyd2001} and \citet{Shen2011}. Finally, the monograph by \citet{Shizgal2015} specifically focuses on applications in physics.

\subsection{Radial mapping strategies}
\label{sec:mapping}

A critical component of spectral methods on semi-infinite domains is the choice of mapping function between the physical coordinate $r$ and the computational coordinate $\xi$. Standard treatments \citep[e.g.][sec.~8.8.2]{Canuto2007} typically focus on mapping the interval $\xi \in [-1, 1]$ to $r \in [0, \infty)$ or $r \in [0, L]$. However, for the specific problem of galactic dynamics, this approach is often suboptimal. The presence of a supermassive black hole creates a cusp or singularity at $r=0$ that is difficult to resolve with global polynomials. Furthermore, the evaluation of stellar densities at extremely large radii ($r \to \infty$) can lead to floating-point underflow.

To address these issues, I generalize the standard mappings to strictly enforce a finite domain $[r_{\min}, r_{\max}]$, where $r_{\min} > 0$ avoids the central singularity and $r_{\max}$ is sufficiently large to approximate the asymptotic regime (e.g. $r_{\max} = \max\{\sigma_{\rm MGE}\}$, or $r_{\max} = 4R_{\rm e}$). I achieve this via a two-step transformation. First, the spectral coordinate $\xi \in [-1, 1]$ is mapped linearly to an intermediate variable $u \in [u_{\min}, u_{\max}]$:
\begin{equation}
    u(\xi) = \frac{u_{\max} - u_{\min}}{2}\xi + \frac{u_{\max} + u_{\min}}{2}.
\end{equation}
Second, the physical radius is defined as $r = f(u)$, where $f$ determines the clustering of grid points. I have implemented and tested five specific mappings, which represent the finite-interval generalizations of the classical semi-infinite stretchings described by \citet{Canuto2007}:

\begin{enumerate}
    \item \textbf{Logarithmic mapping:} $u = \ln r$. This corresponds to the \textit{Truncated Exponential Mapping} \citep[eq.~8.8.14]{Canuto2007}. The nodes are distributed uniformly in $\ln r$.
    \item \textbf{Algebraic mapping:} $u = r/(r+L)$. This corresponds to the \textit{Truncated Algebraic Mapping} \citep[eq.~8.8.13]{Canuto2007}. It maps the semi-infinite domain to a finite interval using a rational function, providing a node distribution that decays as $1/r^2$ at large radii. For this and subsequent mappings, I found good results using as scale parameter the geometric mean of the boundaries, $L = \sqrt{r_{\min} r_{\max}}$.
    \item \textbf{Arctan mapping:} $u = \arctan(r/L)$. This is the trigonometric equivalent of the algebraic mapping, sharing the same asymptotic behaviour and node clustering properties.
    \item \textbf{Asinh mapping:} $u = \operatorname{asinh}(r/L)$. This acts as a hybrid mapping that behaves linearly for $r \ll L$ and logarithmically for $r \gg L$.
    \item \textbf{Linear mapping:} $u = r$. This corresponds to \textit{Domain Truncation} \citep[eq.~8.8.12]{Canuto2007}. It distributes nodes uniformly in the radius $r$.
\end{enumerate}

In extensive tests against the analytic solutions presented in \cref{sec:tests}, I found that the \emph{logarithmic mapping} consistently yielded the best results, producing the smallest relative residuals. The algebraic and arctan mappings performed nearly as well, exhibiting similar exponential convergence properties, provided the scale $L$ was chosen appropriately. The asinh mapping performed poorly, unless the scale was chosen to be extremely small ($L \lesssim r_{\min}$), in which case the mapping effectively reduced to the logarithmic one. Finally, the linear mapping performed remarkably poorly, failing to resolve the gradients in the inner regions even with high grid resolutions.

The superior performance of the logarithmic mapping is not accidental but is a direct consequence of the mathematical structure of the master equation derived in \cref{sec:reduction}. As shown in \cref{eq:final_pde}, the radial differential operator appears explicitly as $\partial / \partial \ln r$. Furthermore, the decay coefficient $\mathcal{C}(r, \theta)$ defined in \cref{eq:coefficient} depends on the term $\partial \ln \nu / \partial \ln r$.

Consider the fundamental building block of galaxy models: a power-law tracer density $\nu \propto r^{-\gamma}$. In linear coordinates, the density slope varies rapidly as $1/r$. However, in the logarithmic coordinate $x = \ln r$, the gradient term becomes a simple constant:
\begin{equation}
    \frac{\partial \ln \nu}{\partial \ln r} = -\gamma.
\end{equation}
Consequently, for a power-law model, the transformation to logarithmic coordinates converts \cref{eq:final_pde} from a differential equation with variable coefficients (which vary rapidly near the centre) into a system with \emph{constant coefficients} in the radial direction.

This means the equation has uniform numerical difficulty across the domain: the local physics at $r=0.01$ is mathematically identical to that at $r=100$. The logarithmic mapping exploits this scale invariance and distributes spectral resolution uniformly across logarithmic decades of galaxy structure, rather than wasting resolution in the outskirts. Because real-galaxy light profiles are typically locally scale-free (e.g., broken power-laws \citep[e.g.][]{Hernquist1990,Lauer2005,Ferrarese2006acs}) or have slowly varying local slopes (e.g., \citet{Sersic1968} profiles, see \citealt{Kormendy2009}), this mapping naturally provides the most efficient and accurate general representation.

\subsection{Chebyshev spectral collocation}
\label{sec:chebyshev}

Having established the logarithmic mapping $x = \ln r$ as the optimal radial coordinate, I solve the governing \cref{eq:final_pde} using a tensor-product Chebyshev collocation spectral method.

The mapped log-radial domain $x \in [\ln r_{\min}, \ln r_{\max}]$ corresponds to the spectral domain $\xi \in [-1, 1]$ as described in \cref{sec:mapping}. Similarly, the angular domain $\theta \in [0, \pi/2]$ is mapped to $\eta \in [-1, 1]$ using the linear transformation:
\begin{equation}
    \theta(\eta) = \frac{\pi}{4}(\eta + 1).
\end{equation}
The derivative operators transform as: 
\begin{equation}
    \frac{\partial}{\partial \ln r} = \frac{2}{\ln(r_{\max}/r_{\min})} \frac{\partial}{\partial \xi}, \quad \frac{\partial}{\partial \theta} = \frac{4}{\pi} \frac{\partial}{\partial \eta}.
\end{equation}

I discretise the domain using Chebyshev--Gauss--Lobatto nodes. For $N_r$ radial points and $N_\theta$ angular points, the nodes are $\xi_i = -\cos(\pi i / N_r)$ and $\eta_j = -\cos(\pi j / N_\theta)$. The unknown function $\overline{v_r^2} = S(x, \theta)$ is represented by a vector $\mathbf{s}$ of size $N_r N_\theta$, formed by flattening the grid values $S_{i,j} = S(x_i, \theta_j)$. Empirically, I found that a good heuristic for defining the grid dimensions to approximately minimise the errors for a given total number of collocation points is to choose $N_r \approx N_\theta \lg(r_{\max}/r_{\min})$.

Differentiation is performed using the standard Chebyshev differentiation matrix $\mathbf{D}$ \citep[e.g.][eq.~6.3--6.5]{Trefethen2000}. For $N+1$ Chebyshev nodes $y_j = \cos(\pi j/N)$ ($j=0, \dots, N$), the entries of $\mathbf{D}$ are:
\begin{align}
    D_{00} &= \frac{2N^2+1}{6}, \quad D_{NN} = -\frac{2N^2+1}{6}, \nonumber \\
    D_{jj} &= \frac{-y_j}{2(1-y_j^2)}, \quad j=1, \dots, N-1, \\
    D_{ij} &= \frac{c_i}{c_j} \frac{(-1)^{i+j}}{y_i - y_j}, \quad i \neq j, \nonumber
\end{align}
where $c_0 = c_N = 2$ and $c_j = 1$ otherwise.

I construct the 1D differentiation matrices $\mathbf{D}_\xi$ (size $N_r \times N_r$) and $\mathbf{D}_\eta$ (size $N_\theta \times N_\theta$). The 2D differential operators are then constructed using Kronecker products ($\otimes$). Due to the general dependence of $\beta(r, \theta)$, the coefficients are not strictly separable. Consequently, I construct the operators by multiplying the derivative matrices with diagonal matrices representing the variable coefficients. While FFT-based methods or separable optimizations could be used, they are not computationally necessary given the small size of the grids (typically $40 \times 30$) studied here.

Let $\mathbf{I}_r$ and $\mathbf{I}_\theta$ be identity matrices of size $N_r$ and $N_\theta$, respectively. The terms in \cref{eq:final_pde} are discretised as follows:

\begin{enumerate}
    \item \textbf{Radial term:} $-\cos\theta \frac{\partial S}{\partial \ln r}$
    \begin{equation}
        \mathbf{L}_{\rm rad} = \operatorname{diag}(-\cos\theta) \left[\frac{2}{\ln(r_{\max}/r_{\min})}\, \mathbf{D}_\xi \otimes \mathbf{I}_\theta \right].
    \end{equation}
    \item \textbf{Angular term:} $(1-\beta)\sin\theta \frac{\partial S}{\partial \theta}$
    \begin{equation}
        \mathbf{L}_{\rm ang} = \operatorname{diag}[(1-\beta)\sin\theta] \left(\frac{4}{\pi}\, \mathbf{I}_r \otimes \mathbf{D}_\eta \right).
    \end{equation}
    \item \textbf{Decay term:} $\mathcal{C} S$
    \begin{equation}
        \mathbf{L}_{\rm decay} = \operatorname{diag}\left( \mathcal{C} \right).
    \end{equation}
\end{enumerate}

The full discrete operator is $\mathbf{L} = \mathbf{L}_{\rm rad} + \mathbf{L}_{\rm ang} + \mathbf{L}_{\rm decay}$. The PDE is thus reduced to the linear system:
\begin{equation}
    \mathbf{L} \cdot \mathbf{s} = \mathbf{b},
\end{equation}
where $\mathbf{b}$ is the vectorised source term $r (\partial\Phi/\partial z)$. Given the relatively small size of the matrix $\mathbf{L}$ (typically less than $1000\times 1000$), I solve this system using a standard dense linear solver (LAPACK; \citealt{Anderson1999lapack}). The boundary conditions are subsequently incorporated into this system as described in \cref{sec:linear_system}.

\subsection{The boundary conditions}
\label{sec:linear_system}

The partial differential equation in \cref{eq:final_pde} is first-order and linear. As illustrated in \citet[fig.~2]{Cappellari2020}, its characteristic curves originate at the outer boundary at infinity and propagate inward to the equatorial plane at different radii. The solution is therefore uniquely set by the boundary condition at infinity, effectively reducing the problem to initial-value ordinary differential equations (ODEs) integrated along characteristics.

For this reason, I do not impose boundary conditions in $\theta$. I verified that adding symmetry boundary conditions (e.g. $\partial \overline{v_r^2}/\partial \theta = 0$) at $\theta=0$ and $\theta=\pi/2$ does not change the numerical solution, consistent with the system's mathematical structure.

Regarding the radial boundary, the natural physical condition is $\nu \overline{v_r^2} \to 0$ as $r \to \infty$. However, spectral methods generally require either a bounded domain or a mapping to infinity. While a rational function mapping \citep[e.g.][sec.~8.8.2]{Canuto2007} can theoretically cover the semi-infinite domain $[0, \infty)$, as discussed in \cref{sec:mapping}, I found that in practice this leads to poor accuracy. Furthermore, mapping to infinity requires evaluating densities at very low values that can underflow to zero, causing numerical instability.

Instead, I adopt a finite radial domain with a large outer boundary $r_{\max}$ (e.g. $r_{\max} \approx \max\{\sigma_{\text{MGE}}\}$). Experiments revealed that simple homogeneous Dirichlet conditions ($\overline{v_r^2}=0$) at this finite boundary introduce significant truncation errors that saturate the exponential convergence curve and prevent the method from reaching machine precision (\cref{sec:tests}). While this error can be mitigated by increasing $r_{\max}$, doing so inefficiently redistributes grid points away from the galaxy centre where kinematic data are typically concentrated.

The proper solution is to impose a Robin boundary condition \citep[e.g.][sec.~1.4]{Strauss2008} that enforces the correct asymptotic power-law decay of the velocity dispersion. This condition takes the form:
\begin{equation}
    \frac{\partial \ln \overline{v_r^2}}{\partial \ln r} = -\mu \quad \Rightarrow \quad \frac{\partial \overline{v_r^2}}{\partial\ln r} + \mu\, \overline{v_r^2} = 0,
    \label{eq:robin_bc}
\end{equation}
where $\mu$ represents the logarithmic slope of the radial velocity dispersion at the boundary.

I determine $\mu$ from the local properties of the potential and tracer density. The derivation proceeds in two steps. First, in \cref{sec:analytic_powerlaw}, I derive a new, fully analytic solution to the spherically-aligned anisotropic Jeans equations. This solution describes an axisymmetric power-law tracer embedded in a spherical power-law potential, assuming constant anisotropy. This result generalizes previous semi-isotropic solutions \citep[][eq.~13.188]{Ciotti2021} and is of independent interest. It demonstrates that the radial and angular dependences of the velocity dispersion are separable, which justifies approximating the dynamics at large radii as a set of independent spherical Jeans equations along each angular ray.

Second, in \cref{sec:robin_correction}, I solve this effective spherical equation assuming a slowly varying power-law density slope. By creating a differential equation for the logarithmic slope $\mu$ and applying an adiabatic approximation (assuming $\partial \mu / \partial r \approx 0$ near the boundary), I derive a correction term that accounts for the curvature of the density profile. This yields an expression for the boundary slope:
\begin{equation}
    \mu \approx \delta + \frac{\gamma'}{\gamma - 2\beta + \delta}.
    \label{eq:mu_approx}
\end{equation}
Here, $\delta \equiv - \partial \ln v_c^2 / \partial \ln r$ is the logarithmic slope of the circular velocity squared (e.g. $\delta=1$ for a Keplerian potential and $\delta=0$ for a constant rotation curve), $\gamma \equiv - \partial \ln \nu / \partial \ln r$ is the tracer density slope, and $\gamma' \equiv \partial \gamma / \partial \ln r$ accounts for the local curvature of the density profile. All quantities required for this correction are already needed and available on the collocation points for the spectral solution, while I compute the $\gamma'$ derivative spectrally at the boundary $(r_{\rm max},\theta_j)$. For real galaxy profiles, where the density slope varies continuously, the second term in \cref{eq:mu_approx} provides a crucial correction that minimizes numerical reflections at the boundary, improving spectral convergence even on truncated grids.

I verified that the full equation, including the curvature-correction $\gamma'$ term, provides a significant reduction of the residuals near the boundary $r_{\max}$. Moreover, all given terms produce measurable improvements. For instance, when testing different anisotropy values, the prediction including $\beta$ is strictly more accurate than ignoring it. In practical applications, the outer boundary $r_{\max}$ is typically placed at large radii (e.g. $r_{\max} \gtrsim 3R_{\rm e}$) to ensure that the line-of-sight integration captures the vast majority of the galaxy's luminosity when computing projected kinematics. However, I tested that this boundary condition is sufficiently accurate to allow for a sub-percent accurate solution even with a boundary $r_{\max}$ as small as $1R_{\rm e}$ for the Sérsic $n=4$ model of \cref{sec:sersic_gnfw_test}.

This boundary condition is implemented directly into the linear system $\mathbf{L} \cdot \mathbf{s} = \mathbf{b}$. I identify the row indices $k$ in the global system vector corresponding to the outer radial boundary nodes ($r = r_{\max}$). For these rows, I replace the standard PDE discretisation in the matrix $\mathbf{L}$ and the source vector $\mathbf{b}$ with the discrete form of \cref{eq:robin_bc}:
\begin{equation}
    \mathbf{L}_{k, :} \leftarrow \left[ \mathbf{D}_{\ln r} \right]_{k, :} + \mu \mathbf{I}_{k, :}, \quad \quad \mathbf{b}_k \leftarrow 0,
\end{equation}
where $[\mathbf{D}_{\ln r}]_{k, :}$ represents the $k$-th row of the logarithmic differentiation matrix (including the mapping metric) expanded into the full 2D operator size, and $\mathbf{I}_{k, :}$ is the $k$-th row of the identity matrix. This enforces the logarithmic slope $\mu$ at the boundary while allowing the value of $S$ to float to satisfy the Jeans solution.

\subsection{Recovering the azimuthal velocity dispersion}
\label{sec:recover_vphi}

Once $\overline{v_r^2}$ is determined, the azimuthal second moment $\overline{v_\phi^2}$ is computed using the radial Jeans equation (\cref{eq:v2phi}), which remains valid for a general $\beta(r, \theta)$. All terms are evaluated spectrally with minimal additional computational cost. The radial derivative $\partial \overline{v_r^2} / \partial \ln r$ is obtained by applying the operator $\mathbf{D}_\xi \otimes \mathbf{I}_\theta$ (including the mapping metric) to the solution vector $\mathbf{s}$. The radial potential gradient $\Phi_r$ is computed via the chain rule, $\Phi_r = \Phi_R \sin\theta + \Phi_z \cos\theta$, using cylindrical potential derivatives; here, $\Phi_z$ is reused from the source term of the spectral solver and does not require recomputation.

\subsection{Modelling the mean rotation}

The solution to the Jeans equations yields the total azimuthal second moment $\overline{v_\phi^2}$. This moment is the sum of the squared mean streaming velocity and the velocity dispersion, $\overline{v_\phi^2} = \overline{v_\phi}^2 + \sigma_\phi^2$, where generally $\sigma_x^2 \equiv \overline{v_x^2} - \overline{v_x}^2$. For the radial and polar coordinates, the mean motion is zero in steady state, implying $\sigma_r^2=\overline{v_r^2}$ and $\sigma_\theta^2=\overline{v_\theta^2}$. However, for the azimuthal component, the tracer distribution and the gravitational potential alone cannot uniquely decompose the total second moment into ordered rotation $\overline{v_\phi}$ and random motions $\sigma_\phi$. In simple terms, for any equilibrium model, one can always reverse the sense of rotation of an arbitrary set of orbits without affecting the density of the tracer or the potential in which they move. Consequently, the sense of rotation---for example, whether a galaxy is a fast or slow rotator (\citealt{Emsellem2007, Emsellem2011})---contains crucial information on the fossil record of galaxy formation \citep[see review by][] {Cappellari2016}, but does not help in constraining the potential. This implies that, in principle, one has complete freedom to parameterise the tangential anisotropy, subject only to the physical constraint that $\overline{v_\phi^2}$ and $\sigma_\phi^2$ must be non-negative (e.g. \citealt{Wang2021, DeDeo2024, DeDeo2025}; \citealt[Sec.~13.3.2]{Ciotti2021}).

To predict the line-of-sight velocity fields, one must adopt a heuristic decomposition. I describe here the case for a single kinematic component. For multi-component systems, which must be applied for MGE models, the approach is described in \citet[sec.~5.2]{Cappellari2020}. A general and flexible approach is to specify the azimuthal anisotropy explicitly, in analogy with the definition of the meridional anisotropy $\beta = 1 - \sigma_\theta^2/\sigma_r^2$. I define the azimuthal anisotropy parameter as:
\begin{equation}
    \gamma(r, \theta) \equiv 1 - \frac{\sigma_\phi^2}{\sigma_r^2}.
\end{equation}
With this definition, the random azimuthal dispersion is given by $\sigma_\phi^2 = (1-\gamma)\overline{v_r^2}$, and the streaming velocity follows directly from the dispersion relation:
\begin{equation}
    \overline{v_\phi} = \left[ \overline{v_\phi^2} - (1 - \gamma)\overline{v_r^2} \right]^{1/2}.
\end{equation}
This formulation allows for a spatially varying anisotropy profile $\gamma(r, \theta)$, providing significant freedom to model complex kinematic structures.

Alternatively, \citet{Satoh1980} introduced a widely used approach to describe the mean rotation of galaxies with a semi-isotropic ($\sigma_R=\sigma_z$ and $\langle v_R v_z\rangle=0$) velocity ellipsoid. This method assumes that the mean rotation is a fraction $\kappa$ of the rotation of a model with an isotropic ($\sigma_\phi=\sigma_R=\sigma_z$) velocity ellipsoid. \citet{Cappellari2008} generalized this approach to describe the velocity fields of anisotropic fast rotators using high-quality IFS data. Although these axisymmetric galaxies generally possess a vertically flattened velocity ellipsoid \citep[$\sigma_z<\sigma_R$;][]{Cappellari2007,Thomas2009}, their velocity fields can be accurately predicted by assuming an \emph{oblate} velocity ellipsoid ($\sigma_\phi=\sigma_R\neq \sigma_z$) and scaling the rotation by $\kappa$ analogously to Satoh's isotropic formulation.

The Satoh approach is naturally applied for cylindrically-aligned velocity ellipsoids, appropriate for flattened systems. In that case, the only natural way to apply it is as $v_\phi=\kappa(\overline{v_\phi^2}-\overline{v_R^2})^{1/2}$. For a spherically aligned velocity ellipsoid, there are two natural possibilities depending on the shape of the system:

(i) \textit{Tangential isotropy} ($\sigma_\phi = \sigma_\theta$): This choice assumes that the velocity ellipsoid has a circular cross-section in the plane orthogonal to the radius. This option makes more sense for close to spherical systems, as it converges to spherical symmetry in the spherical limit. It corresponds to setting $\gamma = \beta$. The resulting streaming velocity is:
\begin{equation}
    \overline{v_\phi} = \kappa \left[ \overline{v_\phi^2} - (1 - \beta)\overline{v_r^2} \right]^{1/2}.
\end{equation}

(ii) \textit{Meridional isotropy} ($\sigma_\phi = \sigma_r$): Alternatively, one may assume the azimuthal random motions track the radial dispersion. This option approximates the cylindrically aligned anisotropy by having an oblate velocity ellipsoid on the galaxy equatorial plane. This form of the anisotropy has been shown to accurately describe the velocity fields of fast rotators (e.g. \citealt[fig.~5]{Cappellari2008}; \citealt[fig.~10]{Cappellari2016}), which are generally flat systems \citep{Weijmans2014}. It corresponds to setting $\gamma = 0$. The streaming velocity becomes:
\begin{equation}
    \overline{v_\phi} = \kappa \left( \overline{v_\phi^2} - \overline{v_r^2} \right)^{1/2}.
\end{equation}

Empirically, it is found that $\kappa \approx 1$ provides an accurate description of fast-rotator galaxies when high-quality data are available (\citealt[fig.~11]{Cappellari2016}; \citealt[fig.~10]{Zhu2023DynPop1}). This suggests that the condition $\sigma_\phi \approx \sigma_r$, inherent to this parametrization, is not merely a convenient fitting function for the mean rotation field, but captures a fundamental property of the internal dynamics of these systems.

Also, note that the value of $\kappa$ does not have to be constant for the whole galaxy. For example, one can have $\kappa$ that changes sign as a function of position \citep{Negri2014} or one can assign different $\kappa$ to different MGE kinematic components (\citealt[fig.~12]{Cappellari2016}; \citealt{Mitzkus2017}; \citealp{Bevacqua2022}). All these studies used these approaches to model counter-rotating stellar components.

All three parametrizations are implemented in \textsc{JamPy}, allowing the user to switch between definitions depending on the specific kinematic features of the galaxy being modelled.

Finally, I emphasize that these kinematic decompositions are heuristic and lack a fundamental physical justification. Consequently, they should not be used when the primary objective is to constrain the gravitational potential, such as when measuring the mass of a supermassive black hole or dark matter. In these applications, one should strictly avoid fitting the mean velocity $V$ and velocity dispersion $\sigma$ separately. Instead, the analysis must rely solely on the total second moment $V^2_{\rm rms} \equiv \overline{v^2_{\rm los}} \approx V^2 + \sigma^2$, which is the unique quantity constrained by the Jeans equations.

\subsection{Spectral interpolation}
\label{sec:interpolation}

To compare the models with observations, I project the intrinsic moments along the line of sight (LOS). A fundamental advantage of the spectral method over finite-difference or discrete quadrature approaches is that it does not merely provide the solution at discrete grid points. Instead, it yields a global functional form for the solution $S(r, \theta)$ expressed as a high-order polynomial. This allows for the evaluation of the smooth solution and its derivatives at any arbitrary location in the domain with spectral accuracy.

I employ barycentric Lagrange interpolation, which is both numerically stable and efficient. For 1D Chebyshev points of the second kind $x_k$ with weights $w_k = (-1)^k \delta_k$ (where $\delta_k=1/2$ at endpoints and 1 otherwise; \citealt[eq.~5.4]{Berrut2004}), the interpolant is given by \citep[eq.~4.2]{Berrut2004}:
\begin{equation}
    f(x) \approx \frac{\sum_{k} \dfrac{w_k}{x-x_k} f_k}{\sum_{k} \dfrac{w_k}{x-x_k}}.
\end{equation}
For the 2D field $S(r, \theta)$, I perform a tensor-product interpolation. Given a point $(r, \theta)$ along the LOS, it is first mapped to spectral coordinates $(\xi, \eta)$. I then interpolate along the angular direction $\eta$ for each radial node to obtain $S_i(\eta)$, and subsequently interpolate these values along the radial direction $\xi$ to obtain $S(\xi, \eta)$.

While barycentric interpolation is computationally more intensive than local low-order methods, the latter introduce interpolation errors that scale algebraically with grid size, potentially degrading the exponential convergence of the spectral solver. In contrast, barycentric interpolation preserves spectral accuracy, ensuring that the interpolation step does not become the accuracy bottleneck. Therefore, I adopt it as the default. However, for applications where maximum speed is critical and moderate accuracy suffices, bilinear or bicubic interpolation remain valid alternatives.

\subsection{Line-of-sight integration}

The LOS integration proceeds by defining a coordinate system $(x', y', z')$, where $z'$ is the coordinate along the LOS. At each sky-plane pixel $(x', y')$, I integrate the luminosity-weighted second moment $\nu \overline{v_{\rm los}^2}$ (or any other moment) along $z'$. At each integration step, coordinates are transformed to the galaxy frame $(r, \theta)$, and the intrinsic moments $\overline{v_r^2}, \overline{v_\theta^2}, \overline{v_\phi^2}$ are evaluated via interpolation, projected onto the LOS vector, and summed using a quadrature rule. For this step, I employ the general tensor-projection formalism described in \citet[sec.~3]{Cappellari2020}.

In my previous implementation \citep[e.g.][sec.~6.3]{Cappellari2020}, the LOS integral was evaluated over the full domain $(-\infty, \infty)$ using a standard TANH-based quadrature \citep{Schwartz1969tahn_quadrature}. This approach is efficient where the integrand is smooth; however, the presence of a supermassive black hole creates a singularity in the potential and velocity dispersion at the galaxy centre ($r=0$). While adaptive quadrature can handle such features, it is ill-suited for the massive parallelisation required by GPU hardware, which relies on fixed instruction sets across threads. 

To resolve this without sacrificing accuracy or resorting to a prohibitively large number of fixed points, I optimize the integration by splitting the interval at $z'=0$ (the point of closest approach to the galaxy centre). This allows the singularity to be treated as a domain endpoint. I apply the ``mixed rule'' quadrature \citep[Table~4.5.14, third column]{Press2007} to each resulting semi-infinite interval. This rule utilises the transformation $z' = \exp(t - e^{-t})$, which provides double-exponential (DE) convergence \citep{Takahasi1974de_quadrature} at the lower limit ($z'=0$), effectively neutralizing the central singularity. It maintains a single-exponential fall-off at infinity, which is sufficient because the MGE surface brightness decreases exponentially and typically becomes negligible beyond $\approx 3\max(\sigma_{\text{MGE}})$. This tailored quadrature allows for high-accuracy integration, achieving errors below the per cent level with as few as 20 points, even for cuspy profiles. By reducing the necessary evaluation points by an order of magnitude, this optimization ensures that the projection step complements the speed of the spectral solver.

\subsection{Computing gradients of density and potential}

The spectral method presented here is fundamentally general: it solves the governing partial differential equations for any tracer density $\nu$ and gravitational potential $\Phi$, provided their gradients can be evaluated at the collocation nodes. The implementation explicitly allows the user to provide arbitrary software functions to compute these gradients. This flexibility represents a significant departure from previous Jeans Anisotropic Modelling (JAM) implementations, which relied entirely on the Multi-Gaussian Expansion (MGE) formalism for both the potential and the density to enable semi-analytic integration.

A typical application of this new capability involves models where an axisymmetric stellar tracer is embedded in a total gravitational potential not derived from the stellar density. A popular choice for the total density is the generalized NFW \citep[gNFW;][]{Wyithe2001} profile, which I employ in the test described in \cref{sec:sersic_gnfw_test}. This setup is frequently utilized in dynamical studies \citep[e.g.][]{Poci2017, Lu2024DynPop5}. In such cases, the gravitational force of the total density often has a simple analytic form \citep[e.g.][]{Zhao1996}. These analytic gradients can be passed directly to the spectral solver, accelerating the computation and, crucially, avoiding the need to approximate the analytic dark matter profile with a one-dimensional MGE fit.

However, for the stellar component, obtaining an accurate intrinsic density from observations requires deprojection of the surface brightness. For this purpose, the MGE method \citep{Emsellem1994, Cappellari2002mge} remains the preferred tool due to its ability to fit complex profiles accurately and deproject them analytically. Consequently, while the solver is general, its integration with MGE components is of primary practical importance. In the following section, I provide the specific formulas required to implement the spectral solver for models where the density and/or potential are described by an MGE.

\section{Implementation for MGE Models}
\label{sec:mge}

The spectral solver requires the logarithmic gradients of the tracer density and the derivatives of the gravitational potential. For Multi-Gaussian Expansion (MGE) models, these quantities can be computed in two ways. The first approach utilises the explicit analytic derivatives of the Gaussian components. The second approach computes the derivatives numerically using the same spectral differentiation matrices employed by the solver itself. I implemented both options in \textsc{JamPy} and found that the spectral differentiation often yields superior results for complex multi-component models. It effectively smooths over small-scale irregularities in the derivatives that arise from the superposition of many Gaussian components, thereby enhancing the stability of the solution. Consequently, I adopt spectral differentiation to generate the lower-accuracy kinematic maps of \cref{fig:mge_maps} and \cref{fig:jam_comparison}, reserving the analytic formulas for the convergence testing of \cref{fig:satoh_convergence} and \cref{fig:mge_convergence}. For completeness, I provide these explicit analytic expressions below.

\subsection{Tracer density derivatives}
\label{sec:tracer_density}

I assume the tracer number density $\nu$ is the deprojection of a surface brightness parameterised by $N$ Gaussian components \citep[e.g.][sec.~4.2]{Cappellari2020}. In cylindrical coordinates $(R, z)$, the density is given by:
\begin{equation}
    \nu(R, z) = \sum_{k=1}^{N} \nu_{0k} \exp\left[ -\frac{1}{2\sigma_k^2} \left( R^2 + \frac{z^2}{q_k^2} \right) \right],
\end{equation}
where $\nu_{0k}$ is the central density, $\sigma_k$ the dispersion, and $q_k$ the intrinsic flattening of the $k$-th component. In spherical coordinates $(r, \theta)$, substituting $R=r\sin\theta$ and $z=r\cos\theta$, the exponent for each component becomes $-r^2/(2\sigma_k^2) [\sin^2\theta + (\cos^2\theta)/q_k^2]$.

The spectral solver requires the logarithmic derivatives of $\nu$. The radial derivative is:
\begin{equation}
    \frac{\partial \ln \nu}{\partial\ln r} = -\frac{1}{\nu(R, z)} \sum_{k=1}^{N} \frac{\nu_k(R, z)}{\sigma_k^2} \left( R^2 + \frac{z^2}{q_k^2} \right),
\end{equation}
where $\nu_k(R, z)$ represents the density contribution of the $k$-th component. The angular derivative is derived from the term $\partial_\theta [\sin^2\theta + \cos^2\theta/q_k^2] = 2\sin\theta\cos\theta - 2\sin\theta\cos\theta/q_k^2 = \sin(2\theta)(1 - 1/q_k^2)$. Thus:
\begin{equation}
    \frac{\partial \ln \nu}{\partial \theta} = -\frac{R\, z}{\nu(R, z)} \sum_{k=1}^{N} \frac{\nu_k(R, z)}{\sigma_k^2} \left( 1 - \frac{1}{q_k^2} \right).
\end{equation}

\subsection{Gravitational potential and forces}
\label{sec:potential_forces}

The source term of the linear system in \cref{eq:final_pde} depends on the vertical potential derivative $\Phi_z$. I assume the potential is generated by a central supermassive black hole of mass $M_\bullet$ and a mass distribution (stars and dark matter) parameterised by an MGE with $M$ components ($\rho_{0j}$, $\sigma_j$, $q_j$). The total potential is $\Phi = \Phi_\bullet + \Phi_{\rm gal}$.

The black hole contribution is $\Phi_\bullet = -GM_\bullet/r$, with $r=\sqrt{R^2+z^2}$, yielding the vertical derivative:
\begin{equation}
    \frac{\partial \Phi_\bullet}{\partial z} = \frac{GM_\bullet z}{r^3}.
\end{equation}

For the galaxy MGE potential $\Phi_{\rm gal}$, I adopt the integral expression from \citet[eq.~44]{Cappellari2020}. 
To highlight the redundancy in the computation of the two derivatives, I define
\begin{equation}
    \Omega_j(R, z, u) \equiv \frac{\rho_{0j} q_j \exp \left[ -\frac{1}{2\sigma_j^2} \left( \frac{R^2}{1+u} + \frac{z^2}{q_j^2+u} \right) \right]}{(1+u)\sqrt{q_j^2+u}}.
\end{equation}
The potential is then:
\begin{equation}
    \Phi_{\rm gal}(R, z) = -2\pi G \int_0^\infty \sum_{j=1}^{M} \sigma_j^2 \Omega_j(R, z, u) \dd u.
\end{equation}
The vertical derivative required for the solver is:
\begin{equation}
    \frac{\partial \Phi_{\rm gal}}{\partial z} = 2\pi G\, z \int_0^\infty \sum_{j=1}^{M} \frac{\Omega_j(R, z, u)}{q_j^2+u} \dd u.
\end{equation}

As discussed in \citet[sec.~6.2]{Cappellari2020}, I evaluate these integrals using the double-exponential (DE) quadrature method (\citealt{Takahasi1974de_quadrature, Trefethen2014}; \citealt[sec.~4.5]{Press2007}). I apply the change of variable $u = \exp(\frac{\pi}{2} \sinh t)$, which maps $t \in (-\infty, \infty)$ to $u \in (0, \infty)$. In practice, I find that integrating over $t \in (-3.7, 3.7)$ ensures a fractional truncation error below $10^{-13}$. This transformation is exceptionally efficient because it forces the integrand to decay double-exponentially at both ends of the interval, allowing the trapezoidal rule to achieve high precision even when the original integrand decays slowly (as $u^{-5/2}$). I have verified that 60 points in $t$ provide $10^{-4}$ relative precision for all common MGE profiles. The quadrature for all $(R,z)$ coordinates in the grid can be computed with a single vectorized operation, leveraging modern CPU and GPU architectures for optimal performance \citep[see also][]{Wang2025}.

To recover $\overline{v_\phi^2}$ via \cref{eq:v2phi}, the radial gradient $\partial \Phi / \partial r$ is also required. For the black hole, $\partial \Phi_\bullet / \partial r = GM_\bullet / r^2$. For the galaxy, I combine the cylindrical derivatives using the chain rule: $\Phi_r = (\partial \Phi_{\rm gal}/\partial R)\sin\theta + (\partial \Phi_{\rm gal}/\partial z)\cos\theta$. The cylindrical radial derivative is:
\begin{equation}
    \frac{\partial \Phi_{\rm gal}}{\partial R} = 2\pi G\, R \int_0^\infty \sum_{j=1}^{M} \frac{\Omega_j(R, z, u)}{1+u} \dd u.
\end{equation}
The total radial force is thus: 
\begin{equation}
    \frac{\partial \Phi}{\partial r} = \frac{GM_\bullet}{r^2} + \sin\theta \frac{\partial \Phi_{\rm gal}}{\partial R} + \cos\theta \frac{\partial \Phi_{\rm gal}}{\partial z}.
\end{equation}
Both $\partial \Phi_{\rm gal}/\partial R$ and $\partial \Phi_{\rm gal}/\partial z$ are computed using the same DE quadrature scheme, ensuring high accuracy and numerical consistency across the spectral grid.

\section{Validation and Performance}
\label{sec:tests}

In this section, I validate the spectral solver by comparing its results against both existing numerical methods and analytic solutions. The primary focus of these comparisons is on the intrinsic velocity moments ($\overline{v_r^2}$ and $\overline{v_\phi^2}$). This choice is motivated by the fact that the projection onto the observational plane relies on the established formalism and software of \citet[sec.~3]{Cappellari2020}. While I have replaced the original bilinear interpolation with a more precise barycentric spectral scheme (\cref{sec:interpolation}), the underlying geometry of the line-of-sight integration remains unchanged. 

Logically, if the intrinsic moments recovered by the spectral solver are correct, the projected observables will be correct by extension. I have explicitly verified this by comparing projected maps from the spectral JAM method against those from the standard quadrature-based JAM; the two approaches yield high-accuracy agreement across the entire sky plane. By focusing the following tests on the intrinsic 3D moments, I provide a more stringent test of the core mathematical breakthrough of this paper: the numerical solution of the Jeans partial differential equations.

\subsection{Comparison with the analytic isotropic Satoh solution}

\begin{figure}
    \centering
    \includegraphics[width=\columnwidth]{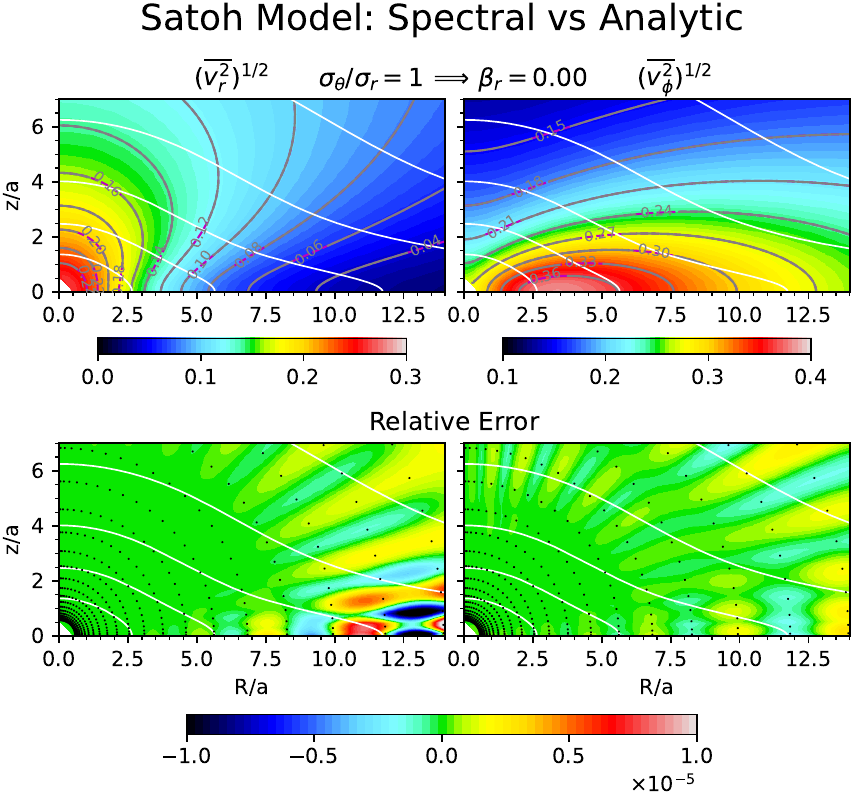}
    \caption{Performance of the spectral JAM solver compared to the exact analytic isotropic Satoh model ($\beta=0$). 
    \textbf{Top panels:} Maps of the intrinsic radial and azimuthal velocity dispersions, $(\overline{v_r^2})^{1/2}$ and $(\overline{v_\phi^2})^{1/2}$, computed on a $32 \times 24$ ($N_r\times N_\theta$) grid with a velocity scale of $\sqrt{G M/a}$. The labeled grey contours show the spectral solution, which closely matches and mostly obscures the analytic solution represented by the dashed magenta contours. The white contours indicate the tracer's isodensity levels, spaced logarithmically by powers of ten from the peak. 
    \textbf{Bottom panels:} Relative percentage error between the spectral and analytic solutions. The black dots mark the collocation points where the spectral approximation exactly satisfies the partial differential equation (PDE). The resulting residuals are essentially structureless and negligible (relative errors $<10^{-5}$) throughout the meridional plane. Note that these contour and point conventions apply to all subsequent kinematic maps.}
    \label{fig:satoh_maps}
\end{figure}

\begin{figure}
    \centering
    \includegraphics[width=\columnwidth]{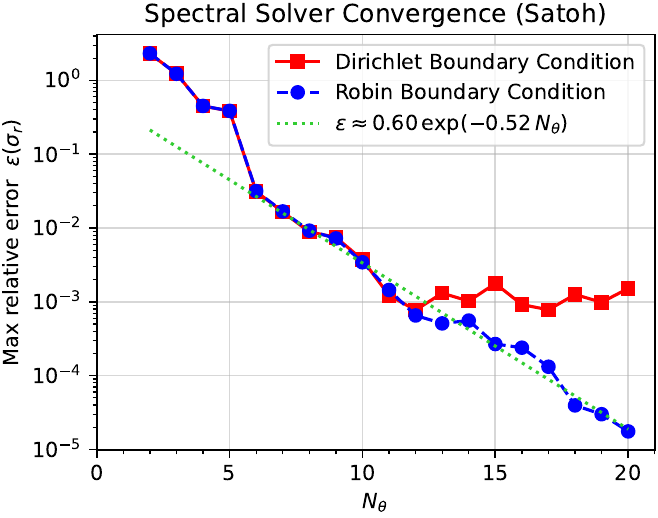}
    \caption{Numerical convergence of the spectral solver as a function of grid resolution. The maximum relative error over the domain $\Omega$ (of size $14a \times 7a$) is defined as $\varepsilon = \max_\Omega |(\sigma_r - \sigma_r^{\mathrm{true}})/\sigma_r^{\mathrm{true}}|$, with the grid dimensions fixed at the ratio $N_r = 2N_\theta$. When a Dirichlet boundary condition ($\overline{v_r^2}=0$) is imposed at the finite boundary of radius $r_{\mathrm{max}}=30a$, the error eventually saturates (red solid line with squares). In contrast, adopting a Robin boundary condition (blue dashed line with circles) restores exponential convergence, allowing the error to decrease steadily toward machine precision. The rapid rate of this convergence is demonstrated by an exponential fit (green dotted line) to the data for $N_\theta \ge 6$. Remarkably, the method achieves an accuracy of $\approx 1\%$ using a grid as coarse as $N_r \times N_\theta = 14 \times 7$.}
    \label{fig:satoh_convergence}
\end{figure}

Fully analytic solutions are essential for the rigorous validation of any new numerical solver. They provide a ground truth that can be evaluated to machine precision and, unlike discretized models, are guaranteed to satisfy the governing equations and boundary conditions exactly. In this first test, I assess the accuracy and convergence properties of the spectral method in isolation, eliminating any uncertainties that might be introduced by the Multi-Gaussian Expansion (MGE) approximation of the density or potential.

I consider the isotropic case of the galaxy model by \citet{Satoh1980}. This model assumes the mass to follow the tracer distribution and possesses a fully analytic solution for the velocity second moments in the semi-isotropic limit, where the velocity ellipsoid has a circular cross-section in the $(R,z)$ meridional plane ($\sigma_R=\sigma_z$ and $\overline{v_R v_z}=0$). This corresponds to the solution provided in \citet[][eqs.~68--69]{Cappellari2020} by setting $\beta_z=0$. To test the solver's pure numerical performance, I provide the \emph{exact} analytic gradients of the Satoh potential and tracer density as inputs to the spectral code, rather than their MGE approximations.

The results are presented in \cref{fig:satoh_maps} and \cref{fig:satoh_convergence}. \cref{fig:satoh_maps} visualises the solution and the relative errors for a representative grid of size $N_r \times N_\theta = 32 \times 24$, which I use in all the subsequent kinematic maps. To facilitate a direct visual comparison with previous implementations, the plot is designed with identical levels and contours as \citet[fig.~5]{Cappellari2020}. The spectral method recovers the intrinsic moments $(\overline{v_r^2})^{1/2}$ and $(\overline{v_\phi^2})^{1/2}$ with remarkable accuracy; the relative errors show the expected structure tracing the distribution of the collocation points, but the deviations remain insignificant and below $10^{-5}$ across the entire meridional plane.

The quantitative convergence is analysed in \cref{fig:satoh_convergence}, which plots the maximum relative error against the number of angular grid points $N_\theta$ (while maintaining $N_r = 2N_\theta$). Unless otherwise specified, throughout this paper, the maximum relative errors and convergence plots always refer to the spatial region shown in the corresponding maps. Near the edges of the computational grid, the relative errors can be larger. However, this is generally irrelevant for practical applications because the galaxy surface brightness typically drops by orders of magnitude at those radii, making the kinematics unobservable.

This test highlights the critical role of the boundary conditions discussed in \cref{sec:linear_system}. In spectral methods, accuracy typically exhibits an initial transient regime, followed by exponential convergence once the solution is adequately resolved. However, as shown by the red squares, when a standard Dirichlet boundary condition ($\overline{v_r^2}=0$) is imposed at the finite outer boundary $r_{\max}$, the convergence quickly saturates around an error floor of $10^{-3}$. This saturation is a direct consequence of the truncation error at the boundary, where the physical solution is small but non-zero.

In contrast, the Robin boundary condition (\cref{eq:robin_bc}) successfully eliminates this saturation (blue circles). The error continues to decrease exponentially toward machine precision limits. The convergence is well-fitted by the exponential $y \approx 0.56\, \exp(-0.51 x)$, demonstrating that the method achieves $\sim 1$ per cent accuracy with as few as $N_\theta=7$ angular points (a $14 \times 7$ grid). This dramatic efficiency gain over traditional grid-based or quadrature methods is a hallmark of properly formulated spectral techniques.

\subsection{Comparison with the analytic anisotropic power-law solution}
\label{sec:powerlaw}

\begin{figure*}
    \centering
    \includegraphics[width=.49\textwidth]{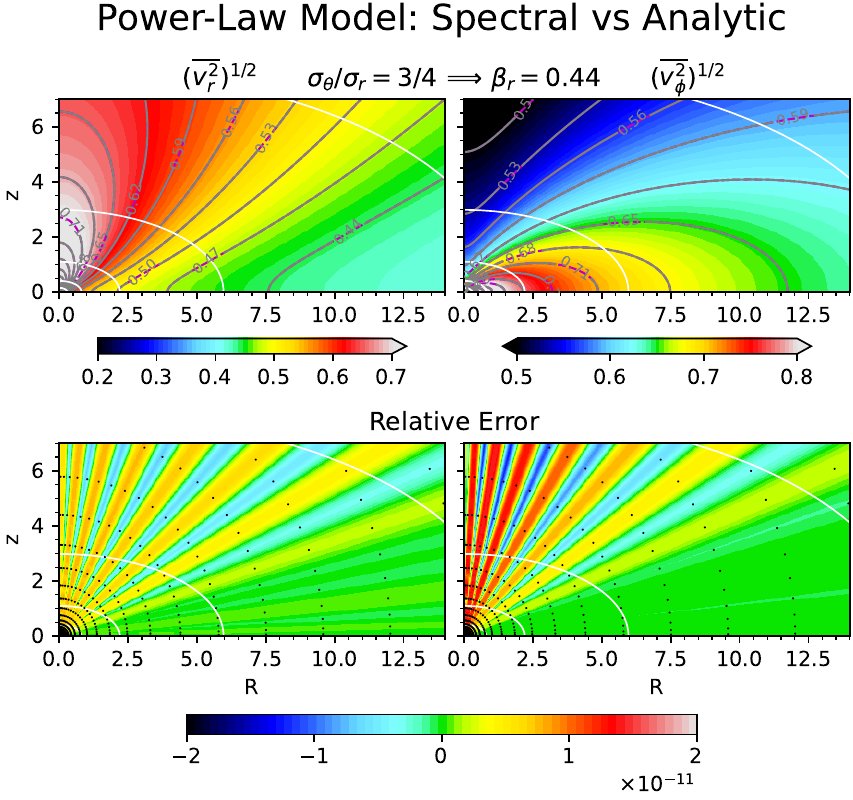}
    \includegraphics[width=.49\textwidth]{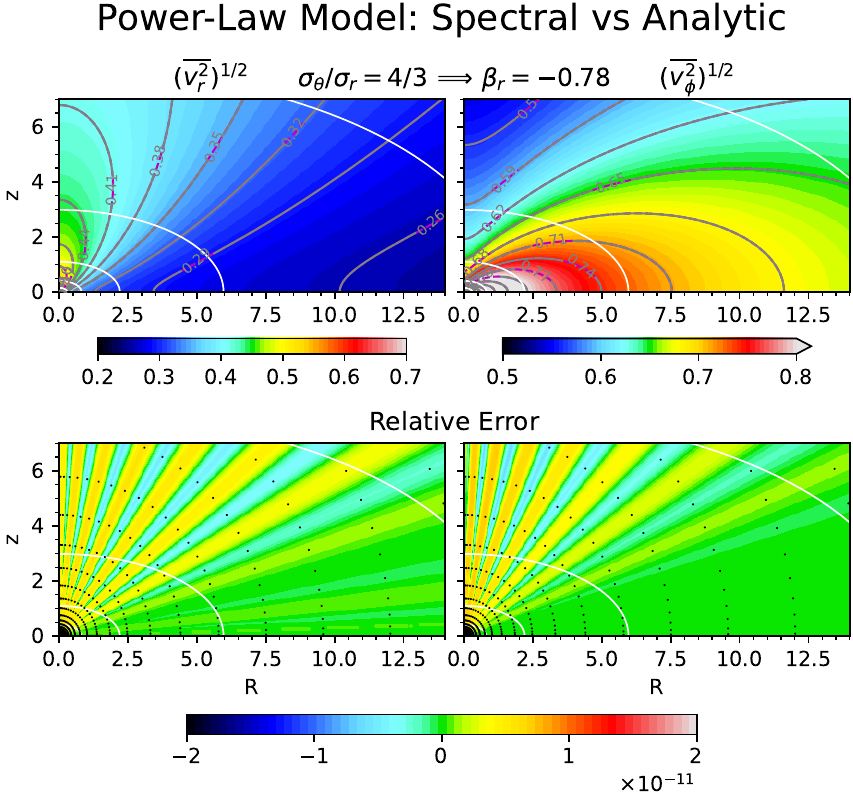}
    \caption{Comparison of the spectral solver with the exact analytic solution for a flattened ($q=0.5$) power-law tracer density $\rho \propto r^{-2.3}$ embedded in a spherical power-law total density $\rho_{\rm tot} \propto r^{-2.2}$.
    \textbf{Left panels:} Radially anisotropic case with $\sigma_\theta/\sigma_r=3/4$.
    \textbf{Right panels:} Tangentially anisotropic case with $\sigma_\theta/\sigma_r=4/3$.
    In both cases, the top rows display the velocity second moment maps, $(\overline{v_r^2})^{1/2}$ and $(\overline{v_\phi^2})^{1/2}$. The spectral solver results are shown as both the background color images and the labeled grey contours, while the analytic solutions are represented by the dashed magenta contours (which are mostly obscured by the grey contours). The contour and point conventions are identical to those described in \cref{fig:satoh_maps}. The velocity scale is normalized to the circular velocity $v_c$ at $R=1$.
    The bottom rows present the relative percentage error between the two solutions, demonstrating that they are nearly identical. The relative error is less than $10^{-11}$, which is effectively machine precision. This result confirms the high accuracy of the spectral solver when applied to anisotropic models.}
    \label{fig:power_law_test}
\end{figure*}

\begin{figure}
    \centering
    \includegraphics[width=\columnwidth]{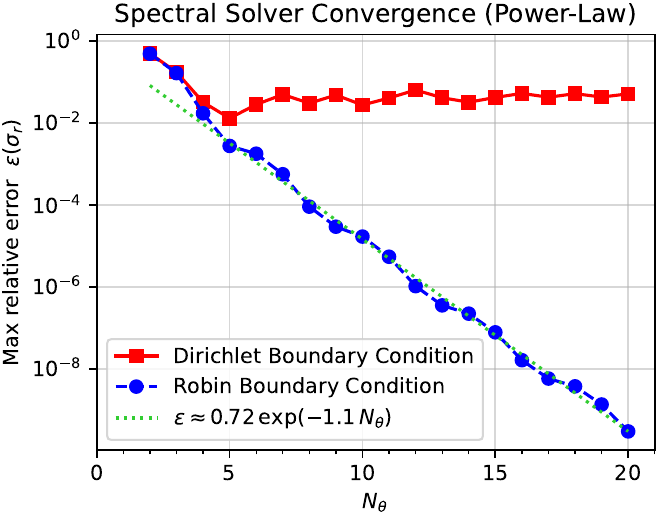}
    \caption{Numerical convergence for the isotropic power-law model ($\beta = 0$). Similar to the behavior shown in \cref{fig:satoh_convergence}, applying a Dirichlet boundary condition at finite radius $r_{\mathrm{max}}=30$ (red solid line) causes the error to saturate above 1 per cent. In contrast, the Robin boundary condition (blue dashed line) maintains exponential convergence down to machine precision as the number of grid points $N_\theta$ increases (with the radial grid size fixed at $N_r=2N_\theta$). Notably, the error of the optimal solution drops to just 1 per cent with a grid as small as $N_\theta=4$. This rapid convergence confirms the optimality of the logarithmic mapping when applied to power-law profiles.}
    \label{fig:power_law_convergence}
\end{figure}

To verify the accuracy of the spectral solver for anisotropic models, I compare it against the exact analytic solution for power-law tracers derived in \cref{sec:analytic_powerlaw}. As in the Satoh test, I use the exact analytic gradients for the density and potential in the spectral solver rather than an MGE approximation. This ensures that the comparison assesses the convergence of the partial differential equation solver itself, rather than the accuracy of the MGE expansion of a power-law.

I construct a model with a flattened spheroidal tracer density ($q=0.5$) scaling as $\rho \propto r^{-2.3}$, embedded in a spherical total mass distribution with density $\rho_{\rm tot} \propto r^{-2.2}$ (implying a circular velocity $v_c^2 \propto r^{-0.2}$). The tracer slope $\gamma\approx2.3$ and total slope $\gamma_{\rm tot}\approx2.2$ used in this test are representative of high-$\sigma$ galaxies, namely massive ETGs (\citealt{Zhu2024DynPop3,Li2024DynPop6}; \citealt[fig.~11]{Cappellari2026}). I test two constant anisotropy values, representing the extreme range of values found in observed galaxies \citep[fig.~10]{Cappellari2026}:
\begin{enumerate}
    \item A radially anisotropic model with $\sigma_\theta/\sigma_r = 3/4$, corresponding to $\beta = 1 - (3/4)^2 \approx 0.44$.
    \item A tangentially anisotropic model with $\sigma_\theta/\sigma_r = 4/3$, corresponding to $\beta = 1 - (4/3)^2 \approx -0.78$.
\end{enumerate}

The results are shown in \cref{fig:power_law_test}, while a qualitative explanation of the observed trends is provided in \cref{sec:gamma_delta_2}. For consistency with the other analytic benchmarks, the convergence analysis (\cref{fig:power_law_convergence}) is performed in the isotropic limit ($\beta=0$). The agreement between the spectral solver and the analytic solution is spectacular, with relative errors of order $10^{-11}$. This effectively machine-precision agreement confirms that the method correctly handles both the anisotropy terms and the singular behaviour of power-law cusps. Notably, a grid with $N_\theta=4$ ($N_r=8$) is already sufficient to reach 1 per cent accuracy.

This test serves as the empirical proof of the optimality of the logarithmic mapping discussed in \cref{sec:mapping}. As visualized in the bottom panels of \cref{fig:power_law_test}, the residuals are strictly constant as a function of radius within each angular sector, exhibiting oscillations only along the angular direction.

This constant error profile confirms that the logarithmic mapping $x = \ln r$ transforms the Jeans operator into one with constant coefficients, making the physics translation-invariant in $\ln r$. Consequently, the truncation error does not vary with radius, producing the flat residual maps observed. Because the radial structure is resolved to essentially machine precision, the convergence error in \cref{fig:power_law_convergence} is dominated entirely by the angular spectral resolution. I verified this by repeating the test with a fixed $N_r=10$ (rather than $N_r=2N_\theta$), obtaining an identical convergence curve; for a power-law, $\approx 10$ radial points suffice to saturate the radial precision.

Any alternative mapping (such as linear or algebraic mappings) would break this symmetry, introducing variable coefficients that degrade accuracy at specific radii---typically near the centre or the boundary. The machine-precision, radially uniform results obtained here demonstrate that the logarithmic mapping provides the mathematically natural coordinate system for these problems. Since real galaxies are well-approximated locally by power-laws, the ability of this mapping to capture such behaviour with a minimal number of basis functions is likely the fundamental reason for the method's remarkable efficiency and overall success.

\subsection{Comparison with the analytic isotropic MGE solution}
\label{sec:sersic_gnfw_test}

\begin{figure}
    \centering
    \includegraphics[width=\columnwidth]{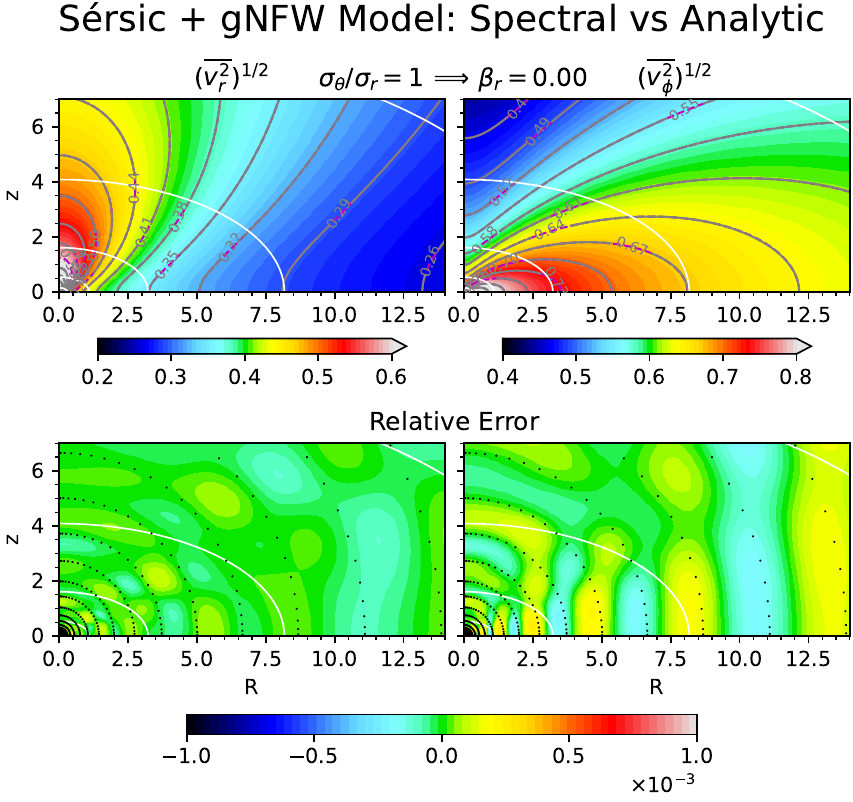}
    \caption{Validation of the spectral solver using a realistic composite galaxy model. The tracer is a Sérsic $n=4$ stellar bulge embedded in a cuspy gNFW total density profile with inner slope $\gamma=2.2$. 
    \textbf{Top panels:} Intrinsic radial and azimuthal velocity dispersion maps $(\overline{v_r^2})^{1/2}$ and $(\overline{v_\phi^2})^{1/2}$. The velocity scale is $\sqrt{G M/r_s}$. 
    \textbf{Bottom panels:} Relative percentage error between the spectral solution and the exact analytic MGE solution derived in \cref{sec:analytic_jeans_mge}. Despite the high dynamic range and central cusp, the maximum error remains remarkably low ($<0.1$ per cent).}
    \label{fig:mge_maps}
\end{figure}

\begin{figure}
    \centering
    \includegraphics[width=\columnwidth]{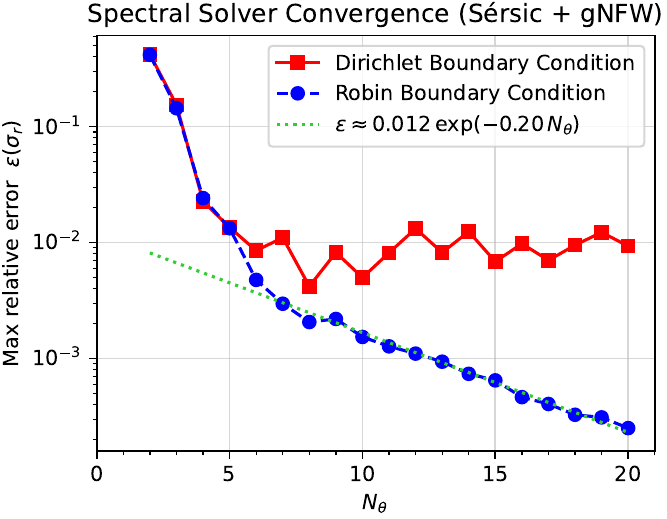}
    \caption{Numerical convergence analysis for the composite Sérsic + gNFW model. The symbols and error definition $\varepsilon$ follow \cref{fig:satoh_convergence}, with the grid dimensions fixed at $N_r=2N_\theta$. Consistent with the Satoh test results, using a Dirichlet boundary condition (red squares) at $r=30=3R_{\rm e}$ results in an accuracy plateau caused by truncation errors. Conversely, the Robin boundary condition (blue circles) restores exponential convergence, allowing the solution to reach a precision of $10^{-4}$. The convergence rate is slightly shallower than in the Satoh case (compare with \cref{fig:satoh_convergence}), which is likely due to the small fluctuations in the MGE model derivatives caused by the superposition of multiple Gaussians.}
    \label{fig:mge_convergence}
\end{figure}

The previous two tests against the Satoh and power-law models validated the spectral solver's core algorithm and its handling of anisotropy using exact analytic gradients. However, both cases described somewhat idealized systems. To evaluate the spectral solver in a regime more representative of massive early-type galaxies (ETGs)---characterized by high central concentrations and significant dark matter fractions---I now construct a more realistic composite model. 

I model the stellar tracer as an axisymmetric oblate distribution with an intrinsic axial ratio $q=0.5$ and a Sérsic $n=4$ profile \citep{deVaucouleurs1948}, typical for massive ETGs \citep[see review by][]{Cappellari2026}. The gravitational potential is generated by a spherical total mass distribution following a generalized NFW (gNFW) profile \citep{Wyithe2001}. I approximate both the tracer and the total density with 20 Gaussians each, using the \textsc{mge\_fit\_1d} function of the \textsc{MgeFit} package\footnote{Current version 6.2 at \url{https://pypi.org/project/mgefit/}} \citep{Cappellari2002mge}. I fit the tracer in the interval $0.1$--$10\,R_{\rm e}$ and the total density in the interval $0.1$--$10\,r_s$, with $R_{\rm e}=10$ (in the arbitrary units used in the plots). I adopt a total density inner slope of $\rho_{\mathrm{tot}} \propto r^{-2.2}$, consistent with observations of massive ETGs out to several effective radii \citep[e.g.][]{Cappellari2015dm, Serra2016, Bellstedt2018}.  

The gNFW break radius is set to $r_s = 10\,R_{\rm e}$. This choice is motivated by the empirical relation between the stellar half-mass radius and halo size, $r_{1/2} \approx 0.015\,R_{200}$ \citep{Kravtsov2013}. Using the approximation $r_{1/2} \approx (4/3)R_{\rm e}$ for a Sérsic $n=4$ profile \citep[Table~2]{Ciotti1991}, this relation can be rewritten as $r_s \approx 89\,R_{\rm e}/c_{200}$. For a representative halo concentration of $c_{200} \approx 9$ \citep[e.g.][]{Dutton2014nfw}, this yields $r_s \approx 10\,R_{\rm e}$. This scale is much larger than the typical field of view and ensures the test probes a regime where the potential has not yet reached its asymptotic Keplerian form.

This setup describes a composite model where the stellar component acts as a tracer in a total potential that I assume to be spherical to provide an exact test case. This configuration is widely used in the dynamical modelling of real galaxies \citep[e.g.][]{Poci2017, Lu2024DynPop5}. Crucially, rather than employing the exact gradients as in the previous sections, here I use the standard MGE implementation described in \cref{sec:mge}. This allows me to validate the ease of use of the method for general applications where analytic gradients are not available. I compare the numerical result against the \emph{analytic} solution for the Jeans equations of a spheroidal MGE tracer in a spherical MGE potential, which I derive in \cref{sec:analytic_jeans_mge}. By using identical MGE coefficients for both the solver and the analytic formulae, I isolate the discretization errors of the PDE solver from any errors associated with MGE fitting.

For all the tests with Satoh, power-law, and Sérsic, I used radial grid boundaries $r_{\min}=\min(\sigma_{\rm MGE})$ and $r_{\max}=30$ (in the units of the plots). This rather arbitrary value of $r_{\max}$, which is about twice the size of the plot and corresponds to just $3\,R_{\rm e}$ for the Sérsic profile, was meant to illustrate the robustness of the Robin condition compared to a Dirichlet one.

The results are presented in \cref{fig:mge_maps} and \cref{fig:mge_convergence}. Despite the high dynamic range of the $n=4$ profile, the spectral method recovers the analytic solution with a maximum relative error $<0.1$ per cent. The convergence properties analysed in \cref{fig:mge_convergence} confirm that the Robin boundary condition (\cref{eq:robin_bc}) restores exponential convergence, achieving a precision of $\sim 10^{-3}$ with a modest $26 \times 13$ grid. This confirms the method is robust for modelling realistic density profiles of galaxies with variable curvature and dark halos.

\subsection{Comparison with numerical anisotropic JAM\textsubscript{sph} solutions}
\label{sec:jam_comparison}

\begin{figure*}
    \centering
    \includegraphics[width=.49\textwidth]{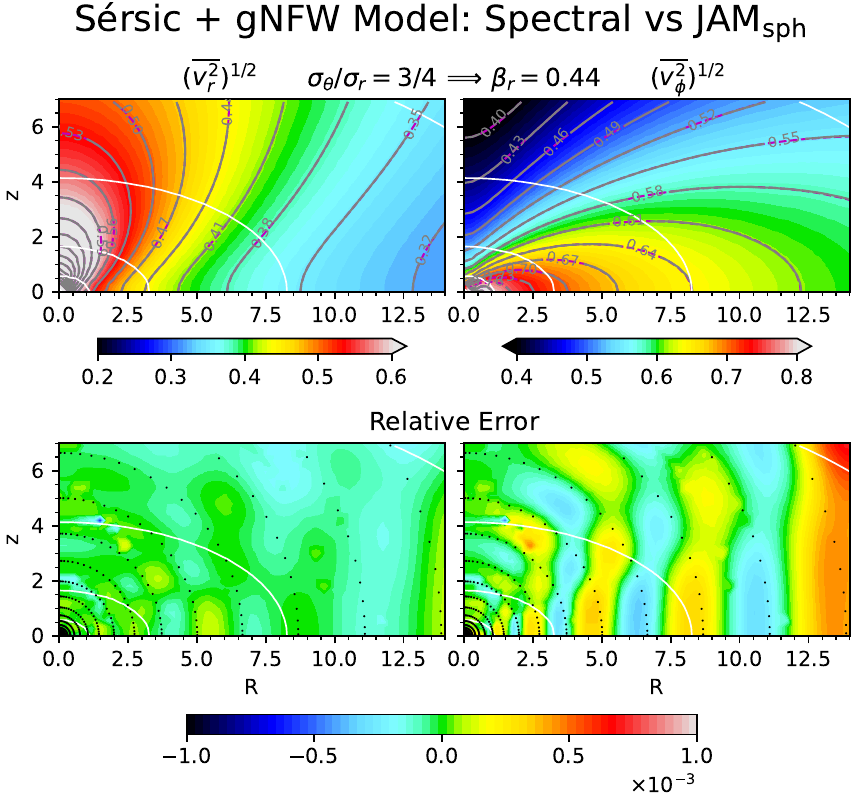}
    \includegraphics[width=.49\textwidth]{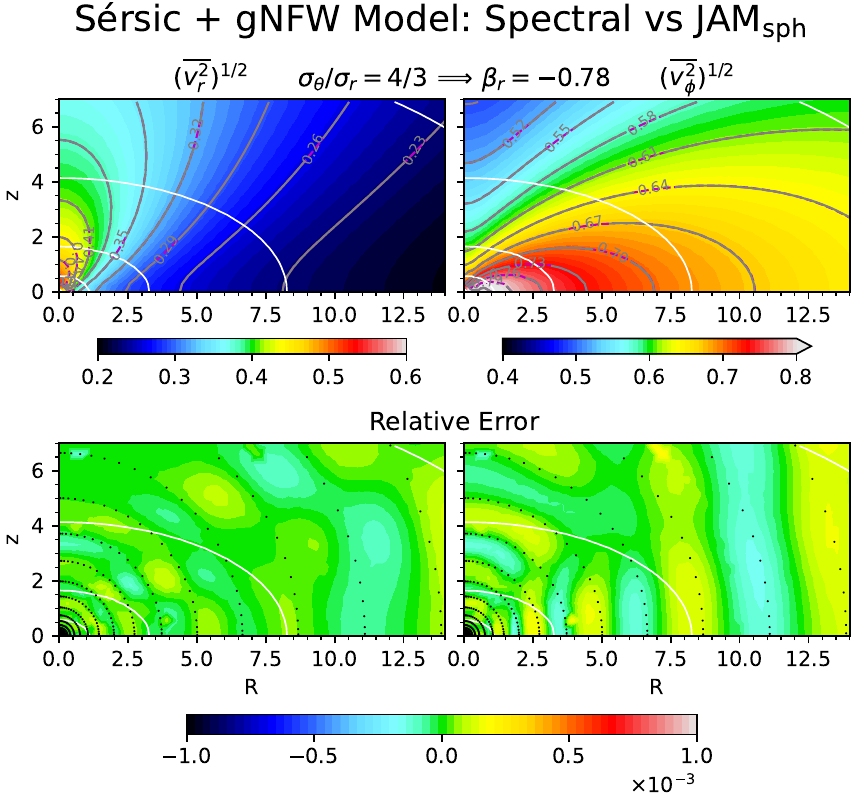}
    \caption{Comparison of the spectral solver against the standard JAM\textsubscript{sph} numerical quadrature for the Sérsic + gNFW model with constant anisotropy.
    \textbf{Left panels:} Radially anisotropic case ($\sigma_\theta/\sigma_r=3/4$).
    \textbf{Right panels:} Tangentially anisotropic case ($\sigma_\theta/\sigma_r=4/3$).
    The top rows display the intrinsic velocity dispersions $(\overline{v_r^2})^{1/2}$ and $(\overline{v_\phi^2})^{1/2}$, while the bottom rows present the relative percentage errors between the two methods. In both anisotropic regimes, the discrepancy remains consistently below $0.1$ per cent, confirming the agreement between the two numerical approaches.}
    \label{fig:jam_comparison}
\end{figure*}

In the general case of a Sérsic + gNFW model with non-zero anisotropy ($\beta \neq 0$), no analytic solution exists for the Jeans equations. To validate the spectral solver in this regime, I compare its performance against the standard public \textsc{JamPy} code \citep{Cappellari2020}. Specifically, I use the JAM\textsubscript{sph} implementation, which utilises a robust numerical scheme based on a direct 2D quadrature of the Jeans equations for a spherically aligned velocity ellipsoid. To ensure that my reference model is more accurate than the spectral solver, I configure JAM\textsubscript{sph} with quadrature with stringent relative error tolerance of $\varepsilon=10^{-4}$.

To probe the solver's stability across different orbital regimes, I test the same two cases of constant anisotropy as in \cref{sec:powerlaw}: $\sigma_\theta/\sigma_r = 3/4$ and $\sigma_\theta/\sigma_r = 4/3$.
The results are presented in \cref{fig:jam_comparison}. In both scenarios, the spectral method produces velocity fields that are virtually indistinguishable from the JAM\textsubscript{sph} results. The relative differences are consistently below $0.1$ per cent in the regions shown. 

Because JAM\textsubscript{sph} is itself a numerical implementation with its own convergence properties, this level of agreement confirms that the spectral solver correctly converges to the same physical solution within the accuracy limits of the reference code. This test successfully validates the implementation of the anisotropy terms and the coupling between the radial and angular components in the spectral formulation. These results demonstrate that the spectral solver can handle the range of orbital configurations typically required for the dynamical modelling of real galaxies.

\subsection{Spectral JAM\textsubscript{sph} with angular anisotropy variation}

\begin{figure*}
    \centering
    \includegraphics[width=.49\textwidth]{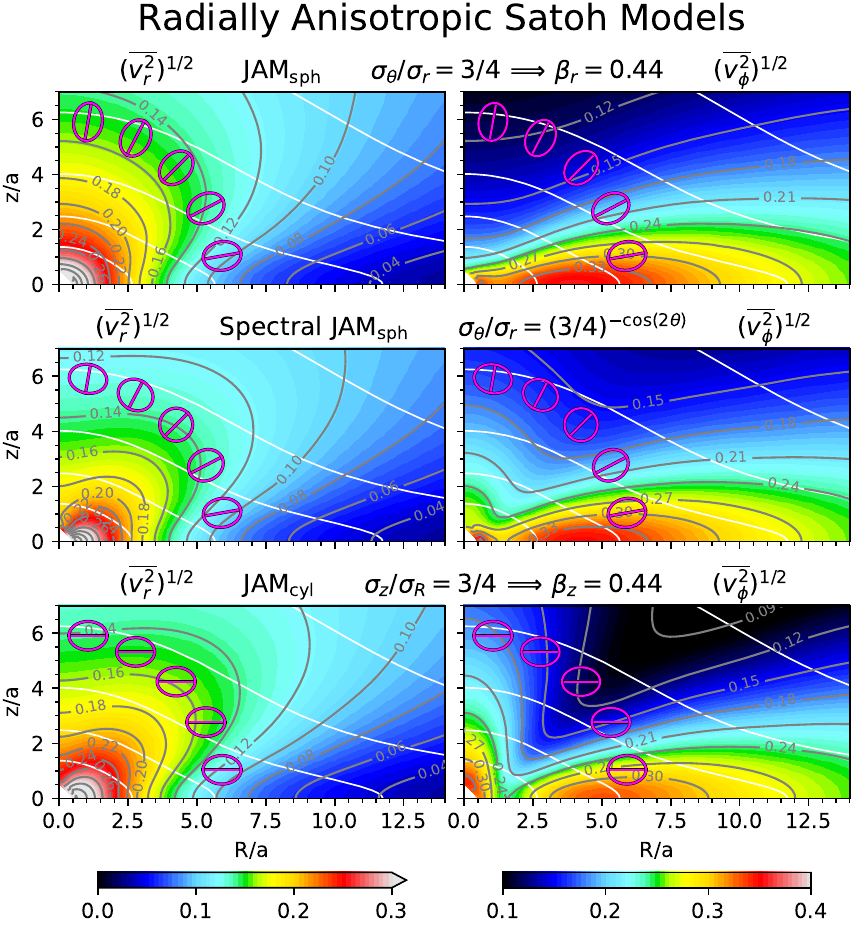}
    \includegraphics[width=.49\textwidth]{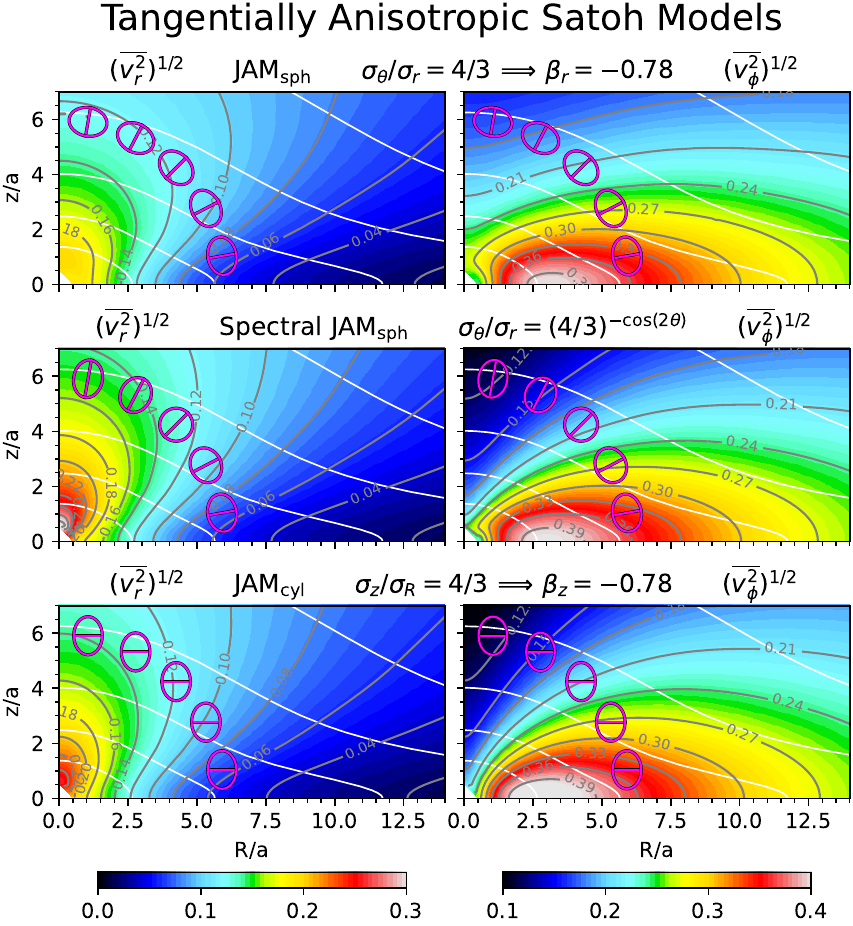}
    \caption{Comparison of velocity second moments under different anisotropy configurations for a Satoh model.
    \textbf{Left panels:} Radially anisotropic models ($\mathcal{R}=3/4$). 
    \textbf{Right panels:} Tangentially anisotropic models ($\mathcal{R}=4/3$).
    \textbf{Top Row:} Standard JAM\textsubscript{sph} with constant spherical anisotropy $\sigma_\theta/\sigma_r=\mathcal{R}$. The velocity ellipsoid is strictly radially oriented (magenta ellipses) and maintains a constant shape everywhere. 
    \textbf{Middle Row:} New Spectral JAM with angularly varying anisotropy $\sigma_\theta/\sigma_r = \mathcal{R}^{-\cos(2\theta)}$. This configuration maintains the spherical coordinate framework but allows the velocity ellipsoid shape (magenta ellipses) to vary with angle, mimicking the behavior observed in realistic galaxies.
    \textbf{Bottom Row:} Standard JAM$_{\rm cyl}$ with constant cylindrical anisotropy $\sigma_z/\sigma_R=\mathcal{R}$. The velocity ellipsoid (magenta ellipses) is cylindrically oriented.
    Note that the Spectral models (middle) successfully reproduce the global kinematic features of the widely-used Cylindrical models (bottom), which are known to approximate fast-rotator galaxies well. Crucially, however, the Spectral models avoid unphysical artifacts, such as the sharp drop in $\overline{v_\phi^2}$ at intermediate angles seen in the bottom-left panel (dark diagonal valley), thus providing a more physically robust solution.}
    \label{fig:anisotropy_variation} 
\end{figure*}

A key novelty of the spectral method presented here is its ability to handle a completely general anisotropy distribution $\beta(r, \theta)$ without compromising computational efficiency. This capability allows for models that are far more physically realistic than those limited by standard assumptions of constant anisotropy or rigid geometric alignments (spherical versus cylindrical).

This flexibility is illustrated in \cref{fig:anisotropy_variation}, where I use the spectral solver to approximate the complex anisotropy structure observed in real galaxies. As detailed in \citet{Cappellari2008} and summarized in their fig.~1, the velocity ellipsoid of fast-rotator early-type galaxies is essentially aligned with the spherical coordinate system $(r, \theta)$ everywhere, consistently with the orbit integration experiment in \cref{sec:justification}. However, the axis of maximum elongation of the ellipsoid is not fixed relative to the radius vector. Instead, the shape of the ellipsoid varies with position: the major axis is radially aligned in the equatorial plane but becomes orthogonal to the radial direction (tangential) near the galaxy symmetry axis (the $z$-axis). This behaviour leads to a global anisotropy that, to first order, resembles a simple flattening of the velocity ellipsoid in the $z$-direction ($\overline{v_z^2} < \overline{v_R^2}$).

The cylindrically aligned JAM models (JAM$_{\rm cyl}$) were originally designed as a simple and efficient heuristic to approximate this variation using a constant anisotropy in cylindrical coordinates ($\beta_z = 1 - \overline{v_z^2}/\overline{v_R^2}$). Here, I demonstrate that the spectral solver can reproduce this realistic behaviour within a spherical alignment framework by adopting an angularly varying anisotropy parameter. I introduce a variation of the form:
\begin{equation}
    \sigma_\theta/\sigma_r = \mathcal{R}^{-\cos(2\theta)} \implies \beta(\theta) = 1 - \mathcal{R}^{-2\cos(2\theta)},
\end{equation}
where $\mathcal{R}$ is the anisotropy ratio parameter. This function ensures that the ratio of tangential to radial dispersion, $\sigma_\theta/\sigma_r$, transitions smoothly from $\mathcal{R}$ at the equatorial plane ($\theta=\pi/2$) to $1/\mathcal{R}$ at the symmetry axis ($\theta=0$). This mimics the variation inherent in cylindrical models where the ratio $\sigma_z/\sigma_R$ is roughly constant.

\cref{fig:anisotropy_variation} compares three scenarios using a Satoh model: 
\begin{enumerate}
    \item Standard JAM\textsubscript{sph} with constant anisotropy (top row).
    \item The new Spectral JAM with the angular variation $\beta(\theta)$ described above (middle row).
    \item JAM$_{\rm cyl}$ with constant $\beta_z$ (bottom row).
\end{enumerate}
I test two ratios: $\mathcal{R}=3/4$ (radially anisotropic at the equator) and $\mathcal{R}=4/3$ (tangentially anisotropic at the equator). The similarity between the JAM$_{\rm cyl}$ models and the new spectral models is striking. This confirms that the success of cylindrical models in fitting fast-rotator kinematics \citep[e.g.][fig.~10]{Cappellari2016} stems from their ability to capture these gross variations in orbital anisotropy.

However, the spectral models offer a crucial improvement. The JAM$_{\rm cyl}$ solution in the radially anisotropic case (bottom-left) exhibits a strong, artificial drop in $\overline{v_\phi^2}$ at intermediate $\theta$ angles (visible as a dark deep diagonal valley in the kinematic map). This is an unphysical artifact of the cylindrical approximation when $\beta_z$ is large and positive. The spectral solution with angular $\beta$ variation (middle-left) reproduces the correct global structure but remains free of such artifacts. Similarly, the spectral result for the tangentially anisotropic case (middle-right) is strikingly similar to the cylindrical counterpart (bottom-right). This test demonstrates the ability of the spectral method to provide physically robust solutions that accurately reflect the underlying dynamical processes, avoiding the singularities introduced by the cylindrical approximation and all without computation speed penalty.

The ability of this spectral solver to handle general $\beta(r, \theta)$ distributions, also makes it a valuable tool for testing and debugging orbit-superposition (Schwarzschild) codes. Specifically, one can construct a Jeans model with the same intrinsic second moments as those fitted by a Schwarzschild code on an $(r, \theta)$ grid. Since the spectral solution is accurate to numerical precision, it can be used to verify that the Schwarzschild code recovers the correct projected moments, thereby quantifying discretization errors in the orbit-integration or weight-solving stages.

\subsection{Execution time benchmarks}

Comparing the execution time of the spectral method against the standard quadrature approach is non-trivial due to their fundamentally different convergence properties. The spectral method yields a global solution with exponential convergence, whereas the standard method relies on independent point-wise quadratures controlled by a tolerance parameter. A rigorous comparison would require tuning the quadrature tolerance to match the \emph{post hoc} accuracy of the spectral solution against an exact analytic solution.

Despite these caveats, a comparative test on identical hardware (an Intel Core i7-1355U CPU) provides a compelling illustration of the efficiency gains. I tested two grid configurations using the standard Python 3.13 implementation, based on the examples that produced \cref{fig:jam_comparison}. For the standard \textsc{JamPy} (version 8.1) benchmarks, I maintained a default relative precision of $\varepsilon=0.01$ for the JAM\textsubscript{sph} quadrature (\texttt{epsrel=0.01}). In both cases, I computed the intrinsic moments ($\overline{v_r^2}$ and $\overline{v_\phi^2}$) for the Sérsic + gNFW model with constant anisotropy $\beta \approx 0.44$ on a polar grid without subsequent interpolation:
 
\begin{enumerate}
    \item \textbf{Standard resolution ($N_r \times N_\theta = 20 \times 10$):} This grid is the default for intrinsic moments in the public \textsc{JamPy} package with semi-isotropic alignment (\texttt{align='sph'}). Comparing the spectral solver against a quadrature evaluation at these points yielded a speedup of $\approx100\times$.
    \item \textbf{High resolution ($N_r \times N_\theta = 32 \times 24$):} On the finer grid used for the error maps in this paper, the spectral method scales more steeply due to the $O(N^3)$ cost of the linear algebra solver, compared to the $O(N)$ linear scaling of independent point-wise quadratures. Nevertheless, the speedup remains a substantial $\approx80\times$.
\end{enumerate}

While the current Python implementations do not yet exploit explicit parallelism, the two methods target different architectures. The quadrature approach was designed for scalar, single-threaded execution, whereas the spectral method consolidates the computation into dense linear algebra operations (matrix-matrix multiplications and system inversions) that are naturally suited to massive parallelization on GPUs and modern multi-core CPUs.

These tests demonstrate that the spectral method provides a dramatic improvement in speed at comparable accuracy. In fact, the Jeans solution becomes so efficient that it is likely no longer the main computational bottleneck of the entire dynamical modelling procedure. This efficiency gain serves as a critical enabler for next-generation modelling pipelines, allowing for the exploration of higher-dimensional parameter spaces.

For this reason, I limit the benchmarks here to the intrinsic moments---the core novelty of this work---rather than characterizing the full modelling execution time, which depends heavily on the implementation details of the other steps like the line-of-sight integration and PSF convolution.

Finally, it is worth noting that speeding up the solver is not the only strategy to accelerate parameter inference. When the parameter space is low-dimensional, one can effectively emulate the model likelihood via interpolation of a grid of pre-computed models \citep[e.g. the analysis of][]{Tdcosmo2025}. Similarly, machine learning techniques are increasingly being used as emulators to bypass direct equation solving, including specifically as a JAM emulator \citep[e.g.][]{Gomer2023, Simon2026neural}. However, these emulator approaches still require a fast and accurate solver to generate training data, and they become less feasible as the dimensionality of the model increases (e.g. when exploring the general anisotropy profiles enabled by this method). The spectral solver presented here thus complements these approaches by providing the fundamental speed and flexibility required for high-dimensional dynamical exploration.

\section{Conclusions}
\label{sec:conclusions} 

In this paper, I have introduced a new spectral solver for the axisymmetric Jeans equations that overcomes the long-standing trade-off between computational speed and physical flexibility in galactic Jeans dynamical modelling. By transforming the partial differential equations into a linear system using Chebyshev collocation, the method achieves a combination of generality and efficiency that was previously unattainable with standard semi-analytic quadrature-based techniques.

The method is built upon three key design choices that ensure both numerical robustness and physical applicability: (i) solving directly for the intrinsic velocity dispersion, rather than the pressure-like second moment, to improve conditioning; (ii) adopting a logarithmic radial coordinate to naturally handle the large dynamic range of galaxies and the singular behaviour of power-law cusps; and (iii) enforcing a rigorous Robin boundary condition to recover the correct asymptotic decay on a finite grid.

The main results and implications of this work are:

\begin{enumerate}
    \item \textbf{General Anisotropy with Spectral Methods:} Unlike standard quadrature schemes that require specific integrability conditions, the spectral method handles completely general anisotropy distributions $\beta(r, \theta)$. This allows for the construction of physically motivated models---such as those with angularly varying anisotropy---that can reproduce the realistic kinematics of fast-rotator galaxies (e.g. mimicking cylindrical alignment) without introducing unphysical artifacts.

    \item \textbf{High-Accuracy Validation:} The solver was validated against three exact analytic benchmarks: the isotropic \citet{Satoh1980} model; a new solution for anisotropic power-law tracers in power-law potentials (\cref{sec:analytic_powerlaw}); and a new analytic solution for an MGE tracer in a spherical MGE potential (\cref{sec:analytic_jeans_mge}). In all tests, the spectral solver recovered the intrinsic moments with negligible errors ($< 0.1$ per cent, reaching machine precision for power-laws), demonstrating the power of spectral methods in high-dynamic-range regimes.

    \item \textbf{Efficiency and Parallelism:} Despite its generality, the spectral method is orders of magnitude faster than traditional high-accuracy quadratures. By reducing the solution of the Jeans equations to standard dense linear algebra operations (matrix-vector products), the algorithm is naturally suited for massive parallelization on GPUs. This efficiency shifts the computational bottleneck of dynamical modelling from the Jeans solver to the line-of-sight integration and PSF convolution, paving the way for the next generation of Bayesian inference pipelines.
\end{enumerate}

This solver enables the exploration of the full solution space allowed by the axisymmetric Jeans equations, facilitating rigorous parameter inference for complex galaxy models. A reference implementation of the spectral JAM method is included in the public \textsc{JamPy} package \citep{Cappellari2008, Cappellari2020}.


\section*{Acknowledgements}

I am grateful to the many users of the \textsc{JamPy} package whose extensive feedback and challenging applications have pushed the limits of the software over the years.
Their ambitious science cases were the primary motivation for developing this new solver.
I thank the referee for an expert and constructive report; their insightful comments substantially improved the presentation, clarity and scope of this paper.

\section*{Data Availability}

No data were generated or analysed in this study.
The code implementing the spectral JAM solver is publicly available in the \textsc{JamPy} package \citep{Cappellari2008, Cappellari2020} at \url{http://pypi.org/project/jampy}.

\appendix
\crefalias{section}{appendix}

\section{Analytic MGE Jeans Solution}
\label{sec:analytic_jeans_mge}
 
To rigorously validate the numerical accuracy of the spectral Jeans Anisotropic Multi-Gaussian Expansion (JAM) method, it is essential to compare its results against exact solutions that share the same underlying physical assumptions. However, analytic solutions to the two-integral or semi-isotropic Jeans equations for non-spherical systems are relatively rare, and general anisotropic solutions are non-existent.

Historically, the Jeans equations, in the isotropic limit, have been solved analytically for several mass-follows-light axisymmetric models---where the gravitational potential is generated by the tracer density distribution. Notable examples include the \citet{Miyamoto1975} model, with its corresponding Jeans solution provided by \citet{Nagai1976}, and the \citet{Satoh1980} model utilized in the main body of this paper. \citet{Smet2015} provides a comprehensive review of such analytical, axisymmetric, isotropic Jeans solutions. More complex isotropic Jeans solutions for composite models, which are capable of describing galaxies embedded in dark matter halos, are even less common. Examples of these include axisymmetric power-law densities within the logarithmic potential of \citet{Binney1981logpot}, or a Keplerian potential, solved by \citet{Evans1993, Evans1994}, and the spheroidal power-law tracers embedded in spherical power-law potentials \citep[][eq.~13.188]{Ciotti2021}. Perhaps the most physically realistic analytic reference currently available is the axisymmetric isotropic Miyamoto--Nagai stellar density embedded in a logarithmic potential dark halo \citep{Smet2015}.

Standard analytic tests, like those using the \citet{Satoh1980} model, require approximating the analytic density with a Multi-Gaussian Expansion (MGE) to validate MGE-based numerical solvers. This introduces fitting errors that can obscure the solver's intrinsic precision. The derivation in this Appendix avoids this by providing a fully analytic solution to the cylindrically-aligned Jeans equations for a flattened axisymmetric MGE tracer embedded in a spherical MGE potential. This formulation uses the exact input parameterization required by the JAM solver, ensuring any discrepancies are due to the numerical scheme, not fitting approximations. While utilizing a spherical potential, this setup remains physically relevant as real galaxy total densities are expected rounder than their stellar tracer distributions, and the MGE's flexibility allows for robust benchmarking across a variety of tracer and density profiles.

\subsection{The vertical velocity dispersion $\overline{v_{z}^2}$}

Following the assumptions of \citet[sec.~3.1]{Cappellari2008}, I consider an axisymmetric system where both the dynamical tracer population and the total mass density are parameterised via the Multi-Gaussian Expansion (MGE). I further assume a cylindrically-aligned velocity ellipsoid. 

For a single axisymmetric Gaussian tracer component $k$, defined by peak density $\nu_{0,k}$, radial dispersion $\sigma_k$, and intrinsic flattening $q_k$, the density distribution is:
\begin{equation}
    \nu_k(R, z) = \nu_{0,k} \exp\left[ -\frac{R^2}{2\sigma_k^2} - \frac{z^2}{2(q_k \sigma_k)^2} \right].
\label{eq:tracer_density}
\end{equation}
The gravitational potential $\Phi$ is generated by a sum of spherical Gaussian mass components $j$, each with total mass $M_j$ and dispersion $\sigma_j$. Imposing the boundary condition $\nu_k \overline{v_{z,k}^2} \to 0$ as $z \to \infty$, the vertical Jeans equation is \citep[eq.~4.223]{Binney2008}:
\begin{equation}
    \nu_k(R, z) \overline{v_{z,k}^2}(R, z) = \int_z^\infty \nu_k(R, z') \frac{\partial \Phi}{\partial z'} \dd z'.
\label{eq:vertical_jeans}
\end{equation}

Defining the Gaussian vertical scale $\sigma_{q,k} \equiv q_k \sigma_k$, the purely radial factor $\exp[-R^2/(2\sigma_k^2)]$ cancels between both sides of \cref{eq:vertical_jeans}. One is left with a one-dimensional integral along $z'$, with $R$ entering only through $r'=\sqrt{R^2+z'^2}$:
\begin{equation}
    \overline{v_{z,k}^2}(R,z)
    = \exp\!\left( \frac{z^2}{2\sigma_{q,k}^2} \right)
    \int_z^\infty \exp\!\left( -\frac{z'^2}{2\sigma_{q,k}^2} \right) \frac{\partial \Phi}{\partial z'} \dd z'.
\end{equation}

For a spherical Gaussian mass component $j$, the vertical force can be written as
\(
\partial \Phi_j/\partial z' = (G M_j(r')/r'^3)\,z'
\),
where $M_j(r')$ is the enclosed mass. Switching variables from $z'$ to the spherical radius $r'$ uses $r'\,\dd r' = z'\,\dd z'$ and maps the limits $z'\in[z,\infty)$ to $r'\in[r,\infty)$, with $r=\sqrt{R^2+z^2}$. The Gaussian kernel transforms into a constant factor times $\exp[-r'^2/(2\sigma_{q,k}^2)]$, so the prefactor and kernel combine into $\exp[r^2/(2\sigma_{q,k}^2)]$. The result depends on $(R,z)$ only through $r$:
\begin{equation}\label{eq:vz2_integral}
    \overline{v_{z,k}^2}(r) = \exp\left( \frac{r^2}{2\sigma_{q,k}^2} \right) \int_r^\infty \exp\left( -\frac{r'^2}{2\sigma_{q,k}^2} \right) \frac{G M_j(r')}{r'^2} \dd r',
\end{equation}
where the enclosed mass for the spherical Gaussian component $j$ is \citep[eq.~49]{Cappellari2008}:
\begin{equation}\label{eq:mr_gaussian}
    M_j(r) = M_j \left[ \operatorname{erf}\left(\frac{r}{\sqrt{2}\sigma_j}\right) - \sqrt{\frac{2}{\pi}} \frac{r}{\sigma_j} \exp\left(-\frac{r^2}{2\sigma_j^2}\right) \right].
\end{equation}

The integral in \cref{eq:vz2_integral} is the same mathematical object that appears for a \emph{spherical} MGE tracer with isotropic dispersion, with the substitution $\sigma_k\to\sigma_{q,k}$. Evaluating it using standard error-function manipulations and the representation in terms of Owen's $T$ function yields, for a single tracer component $k$ in a single spherical Gaussian potential component $j$:
\begin{equation}
    \overline{v_{z,k}^2}(r) = G M_j\left[\frac{1}{r} \operatorname{erf}\left( \frac{r}{\sqrt{2}\sigma_j} \right) 
    - \frac{\sqrt{8\pi}}{\sigma_{q,k}} \exp\left(\frac{r^2}{2\sigma_{q,k}^2}\right) T\left( \frac{r}{\sigma_{q,k}}, \frac{\sigma_{q,k}}{\sigma_j} \right)\right],
    \label{eq:vz2_analytic}
\end{equation}
where $T(\cdot, \cdot)$ is Owen's $T$ function \citep{Owen1956}. This function is readily available in standard numerical libraries, such as \href{https://docs.scipy.org/doc/scipy/reference/generated/scipy.special.owens_t.html}{scipy.special.owens\_t} in Scipy \citep{Scipy2020}, following the implementation by \citet{Patefield2000}, or the built-in \href{https://reference.wolfram.com/language/ref/OwenT.html}{OwensT} in Mathematica \citep{wolfram2003mathematica}. 

Since the gravitational potential is additive, the velocity second moment for a specific tracer component $k$ corresponds to the sum of the contributions from all $M$ Gaussian components of the potential. To obtain the total velocity moments for the full MGE model, the second moments (either $\overline{v_{z}^2}$ or $\overline{v_{\phi}^2}$) are then calculated as the density-weighted average over all $N$ tracer components:
\begin{equation}
\overline{v^2} = \frac{\sum_{k,j} \nu_k\, \overline{v_{k,j}^2} }{\sum_{k} \nu_k}.
\end{equation}

It is important to note that while the vertical velocity dispersion of an individual Gaussian component $\overline{v_{z,k}^2}$ depends only on $r$, the total dispersion $\overline{v_z^2}$---obtained by summing over all tracer and potential components---will exhibit a full $(R, z)$ dependence due to the varying vertical Gaussian scale $\sigma_{q,k}$ across different tracer components.

\subsection{Asymptotic expression to prevent numerical overflow}

The analytic solution for $\overline{v_{z,k}^2}(r)$ in \cref{eq:vz2_analytic} contains the factor $\exp[r^2/(2\sigma_{q,k}^2)]$, which can overflow numerically when $r \gg \sigma_{q,k}$. A naive point-mass approximation is not appropriate unless $r$ is also large compared to \emph{all} potential scales $\sigma_j$; this is often not the case in composite galaxy models, where $r$ may be large relative to a tracer's $\sigma_{q,k}$ yet still comparable to the dark-halo dispersions $\sigma_j$.

To ensure numerical stability in precisely this regime, I use an alternative representation obtained by substituting \cref{eq:mr_gaussian} into \cref{eq:vz2_integral} and rewriting the result as:
\begin{equation}
\overline{v_{z,k}^2}(r)
= \frac{\sqrt{2}\,G M_j}{\sqrt{\pi}\,\sigma_{q,k}}
\int_0^{\sigma_{q,k}/\sigma_j}
\exp\!\left(-\frac{r^2 t^2}{2\sigma_{q,k}^2}\right)
\frac{t^2}{1+t^2}\,dt.
\label{eq:vz2_integral_stable}
\end{equation}

When $r \gg \sigma_{q,k}$, the exponential strongly suppresses the integrand for $t \gtrsim \sigma_{q,k}/r$, so the integral is controlled by the small-$t$ behaviour. In that limit, one may expand
\begin{equation}
\frac{t^2}{1+t^2} = \sum_{n=1}^{\infty} (-1)^{n-1} t^{2n}.
\end{equation}
Numerical tests confirm that \cref{eq:vz2_analytic} remains accurate and matches direct quadrature until overflow occurs; therefore an asymptotic treatment is only needed in the extreme regime $r \gg \sigma_{q,k}$. Retaining only the leading term ($n=1$) gives a compact approximation:
\begin{equation}
\overline{v_{z,k}^2}(r) \simeq
\frac{G M_j\,\sigma_{q,k}^{2}}{r^{2}}
\left[
\frac{1}{r}\,
\operatorname{erf}\left(\frac{r}{\sqrt{2}\sigma_j}\right)
-
\sqrt{\frac{2}{\pi}}\,
\frac{1}{\sigma_j}\,
\exp\!\left(-\frac{r^{2}}{2\sigma_j^{2}}\right)
\right].
\label{eq:vz2_asymptotic_first_term}
\end{equation}
Corrections are $\mathcal{O}[(\sigma_{q,k}/r)^2]$ and are negligible where overflow becomes an issue. Crucially, \cref{eq:vz2_asymptotic_first_term} remains valid even if $r$ is not large compared to $\sigma_j$, making it robust for composite models.

\subsection{The azimuthal velocity second moment $\overline{v_{\phi}^2}$}

For a cylindrically-aligned model where the radial and vertical second moments are related by $\overline{v_R^2} = b\,\overline{v_z^2}$ or equivalently $\beta_z = 1 - \overline{v_z^2}/\overline{v_R^2}$, the semi-isotropic Jeans equation for the azimuthal second moment \citep[eq.~11]{Cappellari2008} is:
\begin{equation}
    \overline{v_{\phi,k}^2} = b\left[\overline{v_{z,k}^2} + \frac{R}{\nu_k} \frac{\partial (\nu_k \overline{v_{z,k}^2})}{\partial R}\right] + R \frac{\partial \Phi}{\partial R}.
\label{eq:vphi_jeans_full}
\end{equation}

Two simplifications make an analytic evaluation straightforward. First, from \cref{eq:tracer_density} one has
\(
(R/\nu_k)\,(\partial \nu_k/\partial R) = -R^2/\sigma_k^2
\).
Second, because $\overline{v_{z,k}^2}$ depends on $(R,z)$ only through $r$, 
\(
\partial \overline{v_{z,k}^2}/\partial R = (R/r)\,\dd \overline{v_{z,k}^2}/\dd r
\).
Substituting these identities into \cref{eq:vphi_jeans_full} gives:
\begin{equation}\label{eq:vphi_intermediate}
    \overline{v_{\phi,k}^2} = b \overline{v_{z,k}^2} \left( 1 - \frac{R^2}{\sigma_k^2} \right) + R \left( b \frac{\partial \overline{v_{z,k}^2}}{\partial R} + \frac{\partial \Phi}{\partial R} \right).
\end{equation}

To eliminate the remaining derivative, one may use the semi-isotropic axisymmetric Jeans relation for a Gaussian tracer profile in a spherical potential. The vertical Jeans equation $\partial (\nu \overline{v_z^2})/\partial z = -\nu \partial \Phi/\partial z$, for a Gaussian tracer with vertical scale $\sigma_{q,k}$ implies $\partial \nu_k/\partial z = -z/\sigma_{q,k}^2 \nu_k$. Since for a spherical potential $\overline{v_{z,k}^2}$ depends only on $r$, one has $\partial \overline{v_{z,k}^2}/\partial z = (z/r)\,\dd \overline{v_{z,k}^2}/\dd r$. Substituting these into the Jeans equation yields:
\begin{equation}
\frac{\dd \overline{v_{z,k}^2}}{\dd r} = -\frac{\dd \Phi}{\dd r} + \frac{r}{\sigma_{q,k}^2} \overline{v_{z,k}^2}.
\end{equation}
For a spherical potential, the radial gradient is given by $\partial \Phi/\partial R = (R/r)\,\dd \Phi/\dd r$. Substituting the gravitational acceleration $\dd \Phi/\dd r = G M_j(r)/r^2$, with $M_j(r)$ from \cref{eq:mr_gaussian}, into the Jeans equation yields the final analytic result:
\begin{equation}
    \overline{v_{\phi,k}^2}(R, z) = b \overline{v_{z,k}^2}(r) 
    \left[ 1 + \frac{(1 - q_k^2)R^2}{\sigma_{q,k}^2} \right] + (1-b) \frac{G M_j(r) R^2}{r^3}.
    \label{eq:vphi2_analytic}
\end{equation}
In the semi-isotropic limit ($b=1$), the term explicitly dependent on the potential gradient vanishes. However, the azimuthal moment is strictly coupled to the vertical moment through a geometric factor that varies with radius $R$, determined by the tracer's flattening ($q_k$). This derivation highlights that even in a spherical potential, the flattening of the tracer distribution explicitly links the spatial coordinates to the azimuthal velocity field.

\subsection{Contribution of a central black hole}
\label{sec:bh_analytic}

Due to the linearity of the Jeans equations with respect to the gravitational potential $\Phi$, the contribution of a central supermassive black hole of mass $M_\bullet$ to the velocity second moments can be computed independently and added to the contribution from the extended mass distribution.

For a point-mass potential $\Phi_\bullet = -GM_\bullet/r$, the integral for the vertical velocity dispersion (\cref{eq:vz2_integral}) can be solved analytically. For a single Gaussian tracer component $k$, the BH-induced vertical second moment is given by \citep[eq.~88]{Emsellem1994}:
\begin{equation}
    \left[\overline{v_{z,k}^2}\right]_\bullet = \frac{GM_\bullet}{r} \left[ 1 - \sqrt{\frac{\pi}{2}} \frac{r}{\sigma_{q,k}} \operatorname{erfcx}\left( \frac{r}{\sqrt{2}\sigma_{q,k}} \right) \right],
\end{equation}
where $\operatorname{erfcx}(x) = \exp(x^2) \operatorname{erfc}(x)$ is the scaled complementary error function (\href{https://docs.scipy.org/doc/scipy/reference/generated/scipy.special.erfcx.html}{scipy.special.erfcx} in Scipy, \citealt{Scipy2020}). This function is essential for avoiding arithmetic overflow when evaluating the combination $\exp[\dots]\operatorname{erfc}[\dots]$ at large radii.

The corresponding contribution to the azimuthal second moment is obtained by applying the BH radial force, $\partial \Phi_\bullet / \partial R = GM_\bullet R / r^3$, to the general solution in \cref{eq:vphi2_analytic}:
\begin{equation}
    \left[\overline{v_{\phi,k}^2}\right]_\bullet = b \left[\overline{v_{z,k}^2}\right]_\bullet 
    \left[ 1 + \frac{(1 - q_k^2)R^2}{\sigma_{q,k}^2} \right] + (1-b) \frac{GM_\bullet R^2}{r^3}.
\end{equation}
This expression reduces to eq.~(102) of \citet{Emsellem1994} in the isotropic case ($b=1$).

The BH contribution should be added to the contributions from the extended mass components to obtain the total second moments.

\section{Analytic Power-Law Jeans Solution}
\label{sec:analytic_powerlaw}

The dynamical problem of power-law axisymmetric spheroidal tracers embedded in spherical power-law potentials was previously studied by \citet[][eq.~13.188]{Ciotti2021}, who derived analytic Jeans solutions for the isotropic case in the specific potentials of a Singular Isothermal Sphere (SIS) and a point mass (Black Hole). These results were recently utilized by \citet{DeDeo2025} to investigate the effects of cylindrically-aligned anisotropy (``\citet{Cappellari2008} $b$-anisotropy''). The derivation below generalizes those semi-isotropic results to systems with \textit{spherically-aligned} velocity ellipsoids with arbitrary constant anisotropy $\beta$, and to arbitrary power-law slope $\delta$ of the circular velocity.

\subsection{Model Definitions}

I assume the tracer number density $\nu$ is a power-law stratified on oblate spheroids with axis ratio $q$. Defining the elliptical radius $m$ in terms of cylindrical coordinates $(R, z)$ as $m^2 = R^2 + z^2/q^2$, the density is
\begin{equation}
    \nu(m) = \nu_0 \left(\frac{m}{r_0}\right)^{-\gamma}.
\end{equation}
In spherical coordinates $(r, \theta)$, this can be rewritten as:
\begin{equation}
    \nu(r, \theta) = \nu_0 q^\gamma \left(\frac{r}{r_0}\right)^{-\gamma} (1 - \omega)^{-\gamma/2}, \quad \text{where } \omega \equiv (1 - q^2) \sin^2 \theta.
\end{equation}
The system is embedded in a spherical potential $\Phi(r)$ that generates a power-law squared circular velocity profile with slope $\delta$:
\begin{equation}
    v_c^2(r) = v_0^2 \left(\frac{r}{r_0}\right)^{-\delta} \implies \frac{\partial \Phi}{\partial r} = \frac{v_c^2(r)}{r}.
\end{equation}
While a power-law density $\rho \propto r^{-\alpha}$ implies a circular velocity slope $\delta = \alpha-2$ (assuming $\alpha < 3$ for a finite $M(<r)$), the formulation in terms of $v_c^2$ is more general. For example, the Keplerian potential of a finite point mass corresponds to $\delta=1$ (the limit $\alpha \to 3$), while a flat rotation curve corresponds to $\delta=0$ (the singular isothermal sphere, $\alpha=2$).

\subsection{Analytic Solution}

For a constant anisotropy parameter $\beta = 1 - \overline{v_\theta^2}/\overline{v_r^2}$, the solution to the axisymmetric Jeans equation with spherically-aligned velocity ellipsoids (\cref{eq:bacon_gen}) can be obtained by integrating along the characteristic curves \citep[e.g.][eqs.~10]{Cappellari2020}:
\begin{equation}
    \nu\overline{v_r^2}(r, \theta) = \int_r^\infty \left( \frac{r'}{r} \right)^{2\beta} \nu(r', \theta') \frac{\dd \Phi}{\dd r}(r') \, \dd r',
\end{equation}
where the integration is done along the path $\sin\theta' = (r'/r)^{\beta-1}\sin\theta$. By performing the change of variable $x = \omega (r'/r)^{2\beta-2}$, the integral reduces to the Incomplete Beta function form $B_\omega(a, b) \equiv \int_0^\omega x^{a-1}(1-x)^{b-1}\mathrm{d}x$ \citep[\href{https://dlmf.nist.gov/8.17.E1}{eq.~8.17.1}]{Olver2010nist}. The final closed-form solution for the intrinsic radial velocity dispersion is separable in $r$ and $\theta$, with the dispersion radial profile $\overline{v_r^2}(r)$ at any angle following the circular velocity $v^2_c(r)$:
\begin{equation}
    \overline{v_r^2}(r, \theta) = v_c^2(r)\frac{(1 - \omega)^{\gamma/2}}{2 - 2\beta} \, \frac{B_\omega(\zeta, 1 - \gamma/2)}{\omega^\zeta},
    \label{eq:analytic_sol}
\end{equation}
where the parameter $\zeta$ depends on the tracer slope $\gamma$, the circular velocity slope $\delta$, and the anisotropy $\beta$:
\begin{equation}
    \zeta \equiv \frac{\gamma - 2\beta + \delta}{2 - 2\beta}.
    \label{eq:zeta_def}
\end{equation} 
One can verify that this expression recovers the isotropic results ($\beta=0$) with $\delta=0$ (SIS) and $\delta=1$ (Keplerian) derived by \citet[][eqs.~13.188]{Ciotti2021}.

In realistic galaxy models, $\zeta$ and $1-\gamma/2$ can be negative, which computer algebra systems such as \emph{Mathematica} handle directly. In numerical libraries restricted to positive parameters (e.g., \texttt{scipy.special.betainc}), the function can instead be evaluated via standard recurrence relations \citep[\href{https://dlmf.nist.gov/8.17.E20}{eq.~8.17.20}--\href{https://dlmf.nist.gov/8.17.E21}{21}]{Olver2010nist} as in \texttt{jampy.util.betax}.

The azimuthal second moment follows from the general Jeans relation \cref{eq:v2phi}. For a power-law model, the radial derivatives simplify to algebraic terms involving the logarithmic slopes: $\partial \ln \nu / \partial \ln r = -\gamma$ and $\partial \ln \overline{v_r^2} / \partial \ln r = -\delta$. Substituting these, along with the circular velocity $v_c^2 = \partial \Phi / \partial \ln r$, yields the linear relation:
\begin{equation}
    \overline{v_\phi^2}(r, \theta) = (1 + \beta - \gamma - \delta)\, \overline{v_r^2}(r, \theta) + v_c^2(r).
\end{equation}

To my knowledge, this is the simplest fully analytic solution to the \emph{anisotropic} spherically-aligned Jeans equations for a flattened tracer, and is therefore an ideal test case for numerical solvers. It accurately describes dynamics both in galaxy outskirts and in centers where stellar densities follow power laws \citep[e.g.][]{Lauer2005}; the approximation is especially accurate in the Keplerian limit ($\delta=1$) inside a supermassive black hole sphere of influence, where the central point mass dominates and the potential is effectively spherical, exactly matching the model's primary assumption.

\subsection{Projected Kinematics}

A notable general property of this scale-free power-law formulation is that the projected line-of-sight $V_{\rm rms}$ separates into a pure radial scaling and a purely angular factor. Introducing the dimensionless line-of-sight coordinate $s \equiv z'/R'$, so that $z' = R's$, one finds that all dependence on the projected radius $R'$ factors out of both the LOS numerator and denominator. This leaves a net scaling $V_{\rm rms}^2(R',\phi') \propto R'^{-\delta}$ along any fixed sky position angle $\phi'$. Consequently, the logarithmic radial slope of the observed $V_{\rm rms}^2$ profile equals $-\delta$, like for the $v_c^2$, and is independent of the tracer slope $\gamma$ and the (constant) anisotropy $\beta$; those parameters affect only the angular dependence and normalization. In particular, a flat rotation curve ($\delta=0$) implies $V_{\rm rms}$ is independent of $R'$ (so iso-$V_{\rm rms}$ contours are radial rays).

\subsection{The Benchmark Case $\gamma+\delta=2$}
\label{sec:gamma_delta_2}

A physically fundamental benchmark is defined by the limit $\gamma + \delta = 2$, a condition that roughly captures the structure of real massive galaxies (e.g. a flat rotation curve $\delta \approx 0$ with a near-isothermal tracer $\gamma \approx 2$). Because of this, the simple analytic relations derived in this limit provide a very useful intuitive understanding of the qualitative kinematic trends observed in realistic galaxy models (see \cref{sec:sersic_gnfw_test} and \cref{sec:jam_comparison}).

In this idealized power-law regime ($\zeta=1$), the anisotropy dependence vanishes from the Beta function term in \cref{eq:analytic_sol}. This implies that anisotropy acts as a global scaling of $\overline{v_r^2}$ through the factor $1/(2-2\beta) = (\sigma_r/\sigma_\theta)^2/2$, without altering the geometric shape of the radial dispersion field.

In contrast, the transverse second moments $\overline{v_\theta^2}$ and $\overline{v_\phi^2}$ are both independent of anisotropy. This is immediate for $\overline{v_\theta^2} = (1-\beta)\overline{v_r^2}$, where the scaling factor $(1-\beta)$ exactly cancels the intrinsic anisotropy scaling of $\overline{v_r^2}$ derived above. It also follows for the azimuthal component $\overline{v_\phi^2}$, as co-adding the transverse moments yields the exact identity:
\begin{equation}
    \overline{v_\phi^2}(r,\theta) + \overline{v_\theta^2}(r,\theta) = v_c^2(r).
\end{equation}
For models satisfying this condition, the sum of the transverse second moments is entirely determined by the local circular velocity. This sum is formally independent of the anisotropy parameter $\beta$, the spatial flattening of the tracer, and the polar angle $\theta$. Thus, while the radial dispersion $\overline{v_r^2}$ scales directly with $(\sigma_r/\sigma_\theta)^2$, the transverse moments remain invariant under changes in anisotropy.

By decomposing the azimuthal moment into its mean and variance components, $\overline{v_\phi^2} = \overline{v_\phi}^2 + \sigma_\phi^2$, and assuming isotropy in the tangential velocity dispersions ($\sigma_\phi = \sigma_\theta \equiv \sigma$), the identity reduces to the form $2\sigma^2 + \overline{v_\phi}^2 = v_c^2$. This expression generalizes the solution for the non-rotating Singular Isothermal Sphere ($\sigma = v_c/\sqrt{2}$ for $\overline{v_\phi}=0$; e.g., \citealt[eq.~4.104]{Binney2008}). It demonstrates mathematically that, for a given $v_c$ in the $\gamma+\delta=2$ limit, the presence of ordered rotation $\overline{v_\phi}$ implies a strictly corresponding reduction in the tangential velocity dispersion $\sigma$.

For completeness, when $\zeta=1$ the Incomplete Beta function reduces to elementary functions, so both $\overline{v_r^2}$ and $\overline{v_\phi^2}$ admit closed forms without special functions, although these algebraic expressions are omitted here for brevity.

\section{Robin Boundary Condition}
\label{sec:robin_correction}

This section details the outer boundary condition implemented in the spectral method. I first justify the use of a spherical approximation to evaluate the boundary condition independently along each angular ray. I then derive the expression for the Robin slope $\mu$, which incorporates a correction for the local density curvature.

\subsection{Justification via Separability}

The general analytic solution for power-law profiles derived in \cref{eq:analytic_sol} takes the separable form:
\begin{equation}
    \overline{v_r^2}(r, \theta) \equiv S(r, \theta) = v_c^2(r) \times \mathcal{F}(\theta, \beta, \gamma, \delta).
\end{equation}
Crucially, the radial scaling of the velocity dispersion is driven entirely by the circular velocity $v_c^2(r)$, while the anisotropy $\beta$, density slope $\gamma$, and potential slope $\delta$ determine the normalization and angular structure. 

In the asymptotic regime ($r \to \infty$), I assume that the system behaves locally like a power-law model. Consequently, one can treat the dynamics along any fixed radial ray at angle $\theta$ as satisfying the spherical Jeans equation, parameterized by the local instantaneous values of the anisotropy $\beta(r_{\max}, \theta)$ and density slope $\gamma(r_{\max}, \theta)$. This allows determining the boundary condition $\mu(\theta)$ for each angular grid point independently.

\subsection{Derivation of the Density Curvature Correction}

Consider the spherically symmetric radial Jeans equation \citep[eq.~4.215]{Binney2008} in logarithmic coordinates $x = \ln r$. Let $S = \overline{v_r^2}$. Defining local logarithmic slopes $\gamma(x) \equiv -\dd \ln \nu/\dd x$ and $\delta(x) \equiv -\dd \ln v_c^2/\dd x$, and allowing for a non-zero curvature $\gamma' \equiv \dd \gamma / \dd x$, the Jeans equation with constant $\beta$ becomes:
\begin{equation}
    \frac{\dd S}{\dd x} - (\gamma - 2\beta) S = -v_c^2.
\end{equation}
I seek the logarithmic slope of the dispersion, $\mu(x) \equiv - \dd \ln S / \dd x$, to apply as my boundary condition. Rearranging the differential equation to isolate $S$:
\begin{equation}
    S \left[ (\gamma - 2\beta) - \frac{1}{S}\frac{\dd S}{\dd x} \right] = v_c^2 \quad \implies \quad S = \frac{v_c^2}{\gamma - 2\beta + \mu}.
    \label{eq:S_rearranged}
\end{equation}
To find $\mu$, I differentiate the logarithm of \cref{eq:S_rearranged} with respect to $x$:
\begin{equation}
    \frac{\dd \ln S}{\dd x} = \frac{\dd \ln v_c^2}{\dd x} - \frac{\dd}{\dd x} \ln (\gamma - 2\beta + \mu).
\end{equation}
Substituting the definitions $\mu = -\dd \ln S / \dd x$ and $\delta = -\dd \ln v_c^2 / \dd x$:
\begin{equation}
    \mu = \delta + \frac{\gamma' - 2\beta' + \mu'}{\gamma - 2\beta + \mu}.
\end{equation}
I now apply the adiabatic approximation. At large radii, I assume the solution relaxes to a power-law form, implying that the slope $\mu$ becomes constant ($\mu' \approx 0$). I also assume constant anisotropy beyond the boundary ($\beta' = 0$). This yields $\mu = \delta + \gamma'/(\gamma - 2\beta + \mu)$. Substituting the zeroth-order approximation $\mu \approx \delta$ into the denominator yields the adopted Robin boundary condition:
\begin{equation}
    \mu \approx \delta + \frac{\gamma'}{\gamma - 2\beta + \delta}.
    \label{eq:robin_corrected}
\end{equation}
Here, $\delta$ represents the dominant gravitational constraint (e.g. $\delta=1$ for a Keplerian potential), while the second term corrects for the fact that the tracer density profile may be steepening or shallowing ($\gamma' \neq 0$) at the grid boundary.

An alternative derivation of \cref{eq:robin_corrected} follows directly from the exact analytic solution in \cref{eq:analytic_sol}. The radial behaviour of the Incomplete Beta function is dominated by the pole in its first parameter, scaling asymptotically as $B_\omega(\zeta, \cdot) \sim \omega^\zeta/\zeta$. This implies that the amplitude of the velocity dispersion scales inversely with $\zeta$. Substituting $\zeta$ from \cref{eq:zeta_def} reveals the fundamental scaling $\overline{v_r^2} \propto v_c^2/(\gamma - 2\beta + \delta)$ of \cref{eq:S_rearranged}. Logarithmic differentiation of this dominant term, treating the density slope $\gamma(r)$ as a variable while holding the potential slope and anisotropy constant, recovers \cref{eq:robin_corrected} exactly.

\label{lastpage}

\end{document}